\documentclass[aps,pre,twocolumn,superscriptaddress,showpacs]{revtex4}
\usepackage{amssymb,amsmath,epsfig,lscape,color,eucal,graphicx,bm,epsfig}
\def\av#1{\left\langle#1\right\rangle}
\def\di{\partial}
\def\to{\rightarrow}
\def\Ka{K_{\alpha}}
\def\sKa{\sqrt{K_{\alpha}}}

\def\Lh{\frac{L}{2}}

\def\x2th{\av{x^2}_{\textrm{th}}}
\def\dixz{\frac{\di}{\di x_0}}
\def\dixzs{\frac{\di^2}{\di x_0^2}}
\def\al{\alpha}

\def\non{\nonumber}
\def\br{\non\\&}
\def\xbar{\overline{x}}

\def\U{\text{U}}
\def\M{\text{M}}
\def\Lfp{{\cal L}_{\textrm{FP}}}
\def\Lfpb{{\cal L}^{\textrm{(B)}}_{\textrm{FP}}}
\def\Lfpt{{\cal L}^{\textrm{(t)}}_{\textrm{FP}}}

\def\mds{-\frac{\di}{\di s}}

\def\Ubar{\overline{U}}

\begin{document}

\title{A fractional Feynman-Kac equation for weak ergodicity breaking}

\author{Shai Carmi}
\affiliation{Department of Physics \& Advanced Materials and
Nanotechnology Institute, Bar-Ilan University, Ramat Gan 52900,
Israel}
\author{Eli Barkai}
\affiliation{Department of Physics \& Advanced Materials and
Nanotechnology Institute, Bar-Ilan University, Ramat Gan 52900,
Israel}

\begin{abstract}
Continuous-time random walk (CTRW) is a model of anomalous sub-diffusion in which particles are
immobilized for random times between successive jumps. A power-law distribution of the waiting
times, $\psi(\tau)\sim \tau^{-(1+\al)}$, leads to sub-diffusion ($\av{x^2}\sim t^{\al}$) for
$0<\al<1$. In closed systems, the long stagnation periods cause time-averages to divert from the
corresponding ensemble averages, which is a manifestation of weak ergodicity breaking. The
time-average of a general observable $\Ubar=\frac{1}{t}\int_0^tU[x(\tau)]d\tau$ is a functional of
the path and is described by the well known Feynman-Kac equation if the motion is Brownian. Here,
we derive forward and backward fractional Feynman-Kac equations for functionals of CTRW in a
binding potential. We use our equations to study two specific time-averages: the fraction of time
spent by a particle in half box, and the time-average of the particle's position in a harmonic
field. In both cases, we obtain the probability density function of the time-averages for $t\to
\infty$ and the first two moments. Our results show that indeed, both the occupation fraction and
the time-averaged position are random variables even for long-times, except for $\al=1$ when they
are identical to their ensemble averages. Using the fractional Feynman-Kac equation, we also study
the dynamics leading to weak ergodicity breaking, namely the convergence of the fluctuations to
their asymptotic values.

\end{abstract}
\date{\today}


\maketitle

\section{Introduction}

The time-average of an observable $U(x)$ of a diffusing particle is defined as
\begin{equation}
\Ubar=\frac{1}{t}\int_0^t U[x(\tau)]d\tau,
\end{equation}
where $x(t)$ is the particle's trajectory. For Brownian motion in a binding potential $V(x)$ and in
contact with a heat bath, ergodicity leads to
\begin{equation}
\lim_{t\to \infty}\Ubar=\av{U}_{\textrm{th}}=\int_{-\infty}^{\infty}U(x)G_{\textrm{eq}}(x)dx,
\end{equation}
where $G_{\textrm{eq}}(x)=e^{-V(x)/(k_BT)}/Z$ is Boltzmann distribution and $\av{U}_{\textrm{th}}$
is the thermal average. The equality of time- and ensemble averages in ergodic systems is one of
the basic presuppositions of statistical mechanics.

In the last decades it was found that in many systems, the diffusion of particles is anomalously
slow, as characterized by the relation $\av{x^2}\sim t^{\al}$ with $0<\al<1$
\cite{Havlin,Bouchaud,KlafterReview2000,AnomalousTransportBook}. Anomalous sub-diffusion is
commonly modeled as a continuous-time random walk (CTRW): nearest-neighbor hopping on a lattice,
with waiting times between jumps distributed as a power-law with infinite mean
\cite{MontrollWeiss,ScherMontroll}.

For closed systems, the long immobilization periods of CTRW result in deviation of time-averages
from ensemble averages even for long times \cite{BarkaiPRL05,BarkaiJSP06,BarkaiPRL07,BarkaiJSP08}.
Although there are no inaccessible regions in the phase space (i.e., there is no strong ergodicity
breaking), the divergence of the mean waiting time results in some waiting times of the order of
magnitude of the entire experiment. Therefore, a particle does not sample the phase space uniformly
in any single experiment, leading to weak ergodicity breaking \cite{BouchaudWEB}.

Two examples of particularly interesting time-averages, which we study in this paper, are given
below. For a particle in a bounded region, the occupation fraction is defined as
$\lambda=\frac{1}{t}\int_0^t \Theta[x(\tau)]d\tau$, namely, it is the fraction of time spent by the
particle in the positive side of the region \cite{GrebenkovPRE,OccTimeMajumdarPRL}. Generally, the
occupation fraction can be defined for any given subspace. Consider, for example, a particle in a
sample illuminated by a laser, where the particle emits photons only when it is under the laser's
focus. The occupation fraction is proportional to the total emitted light
\cite{OccupationExeperimental,AgmonResidenceMoments}. Next, the time-averaged position of a
particle is defined as $\xbar=\frac{1}{t}\int_0^tx(\tau)d\tau$. Recent advances in single particle
tracking technologies enable the experimental determination of the time-average of the position of
beads in polymer networks \cite{AnomalousBeads1,AnomalousBeads2} and of biological macro-molecules
and small organelles in living cells
\cite{AnomalousDNA,AnomalousTelomere,AnomalousmRNA,AnomalousLipid1}. Since in many physical and
biological systems the diffusion is anomalous, the study of occupation fractions or time-averaged
positions in sub-diffusive processes such as CTRW is of current interest.

Time-averages are closely related to functionals, which are defined as $A=\int_0^tU[x(\tau)]d\tau$
and have many applications in physics, mathematics and other fields \cite{MajumdarReview}. Denote
by $G(x,A,t)$ the joint PDF of finding, at time $t$, the particle at $x$ and the functional at $A$.
The Feynman-Kac equation states that for a free Brownian particle \cite{Kac1949}:
\begin{equation}
\label{Feynman-Kac} \frac{\di}{\di t}G(x,p,t)=K_1\frac{\di^2}{\di x^2}G(x,p,t)-pU(x)G(x,p,t),
\end{equation}
where $G(x,p,t)$ is the Laplace transform $A\to p$ of $G(x,A,t)$ and $K_1$ is the diffusion
coefficient. Recently, we developed a \emph{fractional} Feynman-Kac equation for anomalous
diffusion of free particles \cite{BarkaiPRL09,BarkaiJSP10}. As time-averages are in fact scaled
functionals: $\Ubar=A/t$, a generalized Feynman-Kac equation for anomalous functionals in a binding
field would be invaluable for the study of weak ergodicity breaking. Currently, no such equation
exists and weak ergodicity breaking was investigated only in the $t\to \infty$ limit or using
functional- and potential-specific methods \cite{BarkaiPRL05,BarkaiJSP06,BarkaiPRL07,BarkaiJSP08}.

In this paper, we obtain an equation for functionals of anomalous diffusion in a force field
$F(x)$. The equation takes the following form (reported without derivation in \cite{BarkaiPRL09}):
\begin{align}
\label{forward_eq_intro} &\frac{\di}{\di t}G(x,p,t)=\\&\Ka\left[\frac{\di^2}{\di
x^2}-\frac{\di}{\di x}\frac{F(x)}{k_{B}T}\right]{\cal
D}_t^{1-\alpha}G(x,p,t)-pU(x)G(x,p,t)\nonumber.
\end{align}
The symbol ${\cal D}_t^{1-\alpha}$ is a \emph{fractional substantial derivative}, equal in Laplace
$t \to s$ space to $[s+pU(x)]^{1-\alpha}$ \cite{FriedrichPRL06,FriedrichPRE06}, and $\Ka$ is a
generalized diffusion coefficient. Solving Eq. \eqref{forward_eq_intro} for $G(x,p,t)$, inverting
$p\to A$ and integrating over all $x$ yields $G(A,t)$, the PDF of $A$ at time $t$. Changing
variables $A\to A/t=\Ubar$, one finally comes by $G(\Ubar,t)$, the (time-dependent) PDF of $\Ubar$.
Weak ergodicity breaking can then be determined by looking at the long-times properties of
$G(\Ubar,t)$: if $\Ubar$ is not identically equal to $\av{U}_{\textrm{th}}$ for $t\to \infty$,
ergodicity is broken. Moreover, if $G(\Ubar,t)$ or the moments of $\Ubar$ can be found also for
$t<\infty$, the kinetics of weak ergodicity breaking can be uncovered.

In the rest of the paper, we derive Eq. \eqref{forward_eq_intro} as well as a backward equation and
an equation for time-dependent forces. We then apply our equation to the two examples given above:
the occupation fraction in a box and the time-averaged position in a harmonic potential. In both
cases, we calculate the long-times limit of $G(\Ubar,t)$ and the fluctuations $\av{(\Delta
\Ubar)^2}=\av{\Ubar^2}-\av{\Ubar}^2$. We demonstrate that for sub-diffusion both systems exhibit
weak ergodicity breaking, and that the fluctuations decay as $t^{-\al}$ to their asymptotic limit.
Part of the results for the fluctuations of the time-averaged position were briefly reported in
\cite{BarkaiPRL09}.

\section{Derivation of the fractional equations}

\subsection{The forward equation}

\subsubsection{Continuous-time random walk}

\label{sect_CTRW_def}

In the continuous-time random walk model, a particle is placed on an one-dimensional lattice with
spacing $a$ and is allowed to jump to its nearest neighbors only. The probabilities of jumping left
$L(x)$ and right $R(x)$ depend on $F(x)$, the force at $x$ (see next subsection for derivation of
these probabilities). If $F(x)=0$, then $R(x)=L(x)=1/2$. Waiting times between jump events are
independent identically distributed random variables with PDF $\psi(\tau)$, and are independent of
the external force. The initial position of the particle, $x_0$, is distributed according to
$G_0(x)$. The particle waits in $x_0$ for time $\tau$ drawn from $\psi(\tau)$, and then jumps to
either $x_0+a$ (with probability $R(x)$) or $x_0-a$ (with probability $L(x)$), after which the
process is renewed. We assume that the waiting time PDF scales as
\begin{equation}
\label{eq_pdf_psi} \psi(\tau)\sim \frac{B_{\alpha}}{|\Gamma(-\alpha)|}\tau^{-(1+\alpha)},
\end{equation}
where $0<\al<1$. With this PDF, the mean waiting time is infinite and the process is sub-diffusive:
for $F(x)=0$, $x_0=0$, and for an infinite open system, $\av{x^2}\sim t^{\alpha}$
\cite{BarkaiPRE00}. We also consider the case when the mean waiting time is finite, e.g., an
exponential distribution $\psi(\tau)=e^{-\tau/\av{\tau}}/\av{\tau}$. This leads to normal diffusion
$\av{x^2}\sim t$ and we therefore refer to this case as $\al=1$. For discussion on the effect of an
exponential cutoff on Eq. \eqref{eq_pdf_psi}, see \cite{WaitingTimeCutoff}. Below, we derive the
differential equation that describes the distribution of functionals in the continuum limit of this
model.

\subsubsection{Derivation of the equation}

\label{sect_forward_derivation}

Define $A=\int_0^tU[x(\tau)]d\tau$ and define $G(x,A,t)$ as the joint PDF of $x$ and $A$ at time
$t$. For the particle to be at $(x,A)$ at time $t$, it must have been at $[x,A-\tau U(x)]$ at time
$t-\tau$ when the last jump was made. Let $\chi(x,A,t)dt$ be the probability of the particle to
jump into $(x,A)$ in the time interval $[t,t+dt]$. We have,
\begin{equation}
\label{G_Q} G(x,A,t)=\int_0^tW(\tau)\chi[x,A-\tau U(x),t-\tau]d\tau,
\end{equation}
where $W(\tau)=1-\int_0^{\tau}\psi(\tau')d\tau'$ is the probability for {\it not} moving in a time
interval of length $\tau$.

To calculate $\chi$, note that to {\it arrive to} $(x,A)$ at time $t$, the particle must have
arrived to either $[x-a,A-\tau U(x-a)]$ or $[x+a,A-\tau U(x+a)]$ at time $t-\tau$ when the previous
jump was made. Therefore,
\begin{align}
\label{Q_recursion} &\chi(x,A,t)=G_0(x)\delta(A)\delta(t)\\
&+\int_0^t\psi(\tau)L(x+a)\chi[x+a,A-\tau U(x+a),t-\tau]d\tau \nonumber
\\&+\int_0^t\psi(\tau)R(x-a)\chi[x-a,A-\tau U(x-a),t-\tau]d\tau \nonumber.
\end{align}
The term $G_0(x)\delta(A)\delta(t)$ corresponds to the initial condition, namely that at $t=0$,
$A=0$ and the particle's position is distributed as $G_0(x)$.

Assume that $U(x)\geq 0$ for all $x$ and thus $A\geq 0$ (an assumption we will relax in Section
\ref{sect_specialcases}). Let $\chi(x,p,t)=\int_0^{\infty}e^{-pA}\chi(x,A,t)dA$ be the Laplace
transform $A\to p$ of $\chi(x,A,t)$ (we use along this work the convention that the variables in
parenthesis define the space we are working in). Laplace transforming Eq. \eqref{Q_recursion} from
$A$ to $p$, we find
\begin{align}
\label{Q_recursion_Ap} & \chi(x,p,t)=G_0(x)\delta(t) \\
&+L(x+a)\int_0^t\psi(\tau)e^{-p\tau U(x+a)}\chi(x+a,p,t-\tau)d\tau \nonumber \\
&+R(x-a)\int_0^t\psi(\tau)e^{-p\tau U(x-a)}\chi(x-a,p,t-\tau)d\tau\nonumber.
\end{align}
Laplace transforming Eq. \eqref{Q_recursion_Ap} from $t$ to $s$ using the convolution theorem,
\begin{align}
\label{Q_recursion_ts}
&\chi(x,p,s)=G_0(x) \\ & +L(x+a)\hat\psi[s+pU(x+a)]\chi(x+a,p,s) \nonumber \\
&+R(x-a)\hat\psi[s+pU(x-a)]\chi(x-a,p,s) \nonumber,
\end{align}
where $\hat\psi(s)$ is the Laplace transform of the waiting time PDF. Let
$\chi(k,p,s)=\int_{-\infty}^{\infty}e^{ikx}\chi(x,p,s)dx$ be the Fourier transform $x\to k$ of
$\chi$. Fourier transforming Eq. \eqref{Q_recursion_ts} and changing variables $x\pm a\to x$,
\begin{align}
\label{chi_kps_integral}
&\chi(k,p,s)=\hat{G}_0(k)\\
&+e^{-ika}\int_{-\infty}^{\infty}e^{ikx}L(x)\hat\psi[s+pU(x)]\chi(x,p,s)dx \nonumber \\
&+e^{ika}\int_{-\infty}^{\infty}e^{ikx}R(x)\hat\psi[s+pU(x)]\chi(x,p,s)dx\nonumber,
\end{align}
where $\hat{G}_0(k)$ is the Fourier transform of the initial condition.

We now express $L(x)$ and $R(x)$ in terms of the potential $V(x)$. Assuming the system is coupled
to a heat bath at temperature $T$ and assuming detailed balance, we have
\cite{BarkaiPRE00,BarkaiJSP08}
\begin{equation}
\label{detailed_balance1}
L(x)\exp\left[-\frac{V(x)}{k_{B}T}\right]=R(x-a)\exp\left[-\frac{V(x-a)}{k_{B}T}\right].
\end{equation}
If the lattice spacing $a$ is small we can expand
\begin{equation}
\label{detailed_balance2} \exp\left[-\frac{V(x-a)}{k_{B}T}\right] \approx
\exp\left[-\frac{V(x)}{k_{B}T}\right]\left[1-\frac{aF(x)}{k_{B}T} +{\cal O}(a^2)\right],
\end{equation}
where we used $F(x)=-V'(x)$. Expanding $R(x)$ and $L(x)$ for $\frac{aF(x)}{(k_BT)}\ll 1$, using the
fact that $R(x)=L(x)=1/2$ for $F(x)=0$,
\begin{equation}
\label{detailed_balance3} R(x)\approx \frac{1}{2}\left[1+c\frac{aF(x)}{k_BT}\right]=1-L(x),
\end{equation}
where $c$ is a constant to be determined.
Combining Eqs. \eqref{detailed_balance1}, \eqref{detailed_balance2}, and \eqref{detailed_balance3},
we have, up to first order in $a$
\begin{equation*}
1-c\frac{aF(x)}{k_BT}\approx\left[1-\frac{aF(x)}{k_{B}T}\right]\left[1+c\frac{aF(x)}{k_BT}\right]
\end{equation*}
This gives, again up to first order in $a$, $c=1/2$. We can thus write,
\begin{equation}
\label{left_right_force} R(x)\approx
\frac{1}{2}\left[1+\frac{aF(x)}{2k_{B}T}\right]\;;\;L(x)\approx
\frac{1}{2}\left[1-\frac{aF(x)}{2k_{B}T}\right].
\end{equation}

Substituting Eq. \eqref{left_right_force} in Eq. \eqref{chi_kps_integral}, we obtain,
\begin{align*}
&\chi(k,p,s)\approx\hat{G}_0(k)\\
&+\frac{1}{2}e^{-ika}\int_{-\infty}^{\infty}e^{ikx}\hat\psi[s+pU(x)]\chi(x,p,s)dx \nonumber \\
&-\frac{1}{2}e^{-ika}\int_{-\infty}^{\infty}e^{ikx}\frac{aF(x)}{2k_{B}T}\hat\psi[s+pU(x)]\chi(x,p,s)dx \nonumber \\
&+\frac{1}{2}e^{ika}\int_{-\infty}^{\infty}e^{ikx}\hat\psi[s+pU(x)]\chi(x,p,s)dx\nonumber \\
&+\frac{1}{2}e^{ika}\int_{-\infty}^{\infty}e^{ikx}\frac{aF(x)}{2k_{B}T}\hat\psi[s+pU(x)]\chi(x,p,s)dx\nonumber.
\end{align*}
Applying the Fourier transform identity ${\cal F}\{xf(x)\}=-i\frac{\di}{\di k}f(k)$, the last
equation simplifies to
\begin{align}
\label{Q_recursion_kps}
\chi(k,p,s)\approx\hat{G}_0(k)+&\left[\cos(ka)+i\sin(ka)\frac{aF\left(-i\frac{\di}{\di
k}\right)}{2k_{B}T}\right]\times \nonumber\\ &\hat\psi\left[s+pU\left(-i\frac{\di}{\di
k}\right)\right]\chi(k,p,s).
\end{align}
The symbols $F\left(-i\frac{\di}{\di k}\right)$ and $U\left(-i\frac{\di}{\di k}\right)$ represent
the original functions $F(x)$ and $U(x)$, but with $-i\frac{\di}{\di k}$ as their arguments. Note
that the order of the terms is important: for example, $\cos(ka)$ does not commute with
$\hat\psi\left[s+pU\left(-i\frac{\di}{\di k}\right)\right]$. The formal solution of Eq.
\eqref{Q_recursion_kps} is
\begin{align}
\label{Q_solution}
\chi(k,p,s)\approx&\left\{1-\left[\cos(ka)+i\sin(ka)\frac{aF\left(-i\frac{\di}{\di
k}\right)}{2k_{B}T}\right]\right.\times \nonumber \\
&\left.\hat\psi\left[s+pU\left(-i\frac{\di}{\di k}\right)\right]\right\}^{-1}\hat{G}_0(k).
\end{align}
We next use our expression for $\chi$ to calculate $G(x,A,t)$. Transforming Eq. \eqref{G_Q}
$(x,A,t)\to (k,p,s)$,
\begin{equation}
\label{G_Q_kps} G(k,p,s)= \frac{1-\hat\psi\left[s+pU\left(-i\frac{\di}{\di
k}\right)\right]}{s+pU\left(-i\frac{\di}{\di k}\right)}\chi(k,p,s),
\end{equation}
where we used the fact that $\hat W(s)=[1-\hat\psi(s)]/s$. Substituting Eq. \eqref{Q_solution} into
\eqref{G_Q_kps}, we have
\begin{align}
\label{P_usk} &G(k,p,s)\approx \frac{1-\hat\psi\left[s+pU\left(-i\frac{\di}{\di
k}\right)\right]}{s+pU\left(-i\frac{\di}{\di k}\right)} \times \\
&\left\{1-\left[\cos(ka)+i\sin(ka)\frac{aF\left(-i\frac{\di}{\di
k}\right)}{2k_{B}T}\right]\right.\times \nonumber \\
&\left.\hat\psi\left[s+pU\left(-i\frac{\di}{\di k}\right)\right]\right\}^{-1}\hat{G}_0(k)\nonumber.
\end{align}

To derive a differential equation for $G(x,p,t)$, we use the small $s$ expansion of $\hat\psi(s)$.
For $0<\alpha<1$, where the waiting time PDF is $\psi(\tau) \sim
B_{\alpha}\tau^{-(1+\alpha)}/|\Gamma(-\alpha)|$ (Eq. \eqref{eq_pdf_psi}), the Laplace transform for
small $s$ is \cite{KlafterReview2000}
\begin{equation}
\label{psi_s_small} \hat\psi(s) \approx 1-B_{\alpha}s^{\alpha}.
\end{equation}
The case $\al=1$ is also described by Eq. \eqref{psi_s_small}, if we identify $B_1$ with the mean
waiting time $\av{\tau}$. Substituting Eq. \eqref{psi_s_small} in Eq. \eqref{P_usk}, and using
$\cos(ka) \approx 1-k^2a^2/2$ and $\sin(ka)\approx ka$, we obtain
\begin{align}
\label{G_psk_cont} &G(k,p,s)\approx \left[s+pU\left(-i\frac{\di}{\di k}\right)\right]^{\alpha-1} \times\\
&\left\{\Ka \left[k^2-ik\frac{F\left(-i\frac{\di}{\di
k}\right)}{k_{B}T}\right]+\left[s+pU\left(-i\frac{\di}{\di
k}\right)\right]^{\alpha}\right\}^{-1}\hat{G}_0(k) \nonumber,
\end{align}
where we defined the generalized diffusion coefficient \cite{BarkaiPRE00}
\begin{equation}
\label{gen_diff_coeff} \Ka\equiv \lim_{a^2,B_{\alpha}\to 0}\frac{a^2}{2B_{\alpha}},
\end{equation}
with units $\textrm{m}^2/\textrm{sec}^{\alpha}$. Rearranging the expression in Eq.
\eqref{G_psk_cont},
\begin{align*}
&sG(k,p,s)-\hat{G}_0(k)=-pU\left(-i\frac{\di}{\di
k}\right)G(k,p,s)\\&-\Ka\left[k^2-ik\frac{F\left(-i\frac{\di}{\di
k}\right)}{k_{B}T}\right]\left[s+pU\left(-i\frac{\di}{\di
k}\right)\right]^{1-\alpha}G(k,p,s).\nonumber
\end{align*}
Inverting $k\to x, s\to t$, we finally obtain our fractional Feynman-Kac equation:
\begin{equation}
\label{forward_eq_derive}
\frac{\di}{\di t}G(x,p,t)=\Ka\Lfp{\cal D}_t^{1-\alpha}G(x,p,t)-pU(x)G(x,p,t).
\end{equation}
The symbol $\Lfp$ represents the Fokker-Planck operator,
\begin{equation}
\label{fokker-planck_def} \Lfp=\frac{\di^2}{\di x^2}-\frac{\di}{\di x}\frac{F(x)}{k_{B}T},
\end{equation}
and the initial condition is $G(x,A,t=0)=G_0(x)\delta(A)$, or $G(x,p,t=0)=G_0(x)$. The symbol
${\cal D}_t^{1-\alpha}$ represents the fractional substantial derivative operator introduced in
\cite{FriedrichPRL06,MetzlerLevyWalk}:
\begin{equation}
\label{substantial_def_s} {\cal L}\left\{{\cal
D}_t^{1-\alpha}G(x,p,t)\right\}=[s+pU(x)]^{1-\alpha}G(x,p,s),
\end{equation}
where ${\cal L}\{f(t)\}=\int_0^{\infty}e^{-st}f(t)dt$ is the Laplace transform $t\to s$. In $t$
space,
\begin{align}
\label{substantial_def} &{\cal D}_t^{1-\alpha}G(x,p,t)=\\
&\frac{1}{\Gamma(\alpha)}\left[\frac{\di}{\di
t}+pU(x)\right]\int_0^t\frac{e^{-(t-\tau)pU(x)}}{(t-\tau)^{1-\alpha}}G(x,p,\tau)d\tau \nonumber.
\end{align}
Thus, due to the long waiting times, the evolution of $G(x,p,t)$ is non-Markovian and depends on
the entire history.

\subsubsection{Special cases and extensions}

\label{sect_specialcases}

\emph{Normal diffusion.---} For $\al=1$, or normal diffusion, the fractional substantial derivative
equals unity and we have
\begin{equation}
\frac{\di}{\di t}G(x,p,t)=K_1\Lfp G(x,p,t)-pU(x)G(x,p,t).
\end{equation}
This is simply the (integer) Feynman-Kac equation \eqref{Feynman-Kac}, extended to a general force
field $F(x)$.

\emph{The fractional Fokker-Planck equation.---} For $p=0$, $G(x,p=0,t)=\int_0^{\infty}G(x,A,t)dA$
reduces to $G(x,t)$, the marginal PDF of finding the particle at $x$ at time $t$ regardless of the
value of $A$. Correspondingly, Eq. \eqref{forward_eq_derive} reduces to the fractional
Fokker-Planck equation \cite{BarkaiPRL99,BarkaiPRE00,BarkaiPRE01}:
\begin{equation}
\label{fractional_FP} \frac{\di}{\di t}G(x,t)=\Ka\Lfp{\cal D}_{\textrm{RL},t}^{1-\alpha}G(x,t),
\end{equation}
where ${\cal D}_{\textrm{RL},t}^{1-\alpha}=\left.{\cal D}_t^{1-\alpha}\right|_{p=0}$ is the
Riemann-Liouville fractional derivative operator. In Laplace $s$ space, ${\cal
D}_{\textrm{RL},t}^{1-\alpha}G(x,s)=s^{1-\al}G(x,s)$.

\emph{Free particle.---} For $F(x)=0$, $\Lfp=\frac{\di^2}{\di x^2}$. Several applications of this
special case were treated in \cite{BarkaiJSP10}.

\emph{A general functional.---} When the functional is not necessarily positive, the Laplace
transform $A\to p$ is replaced by a Fourier transform
$G(x,p,t)=\int_{-\infty}^{\infty}e^{ipA}G(x,A,t)dA$. The fractional Feynman-Kac equation looks like
\eqref{forward_eq_derive}, but with $p$ replaced by $-ip$,
\begin{equation}
\label{forward_eq_fourier} \frac{\di}{\di t}G(x,p,t)=\Ka\Lfp{\cal
D}_t^{1-\alpha}G(x,p,t)+ipU(x)G(x,p,t),
\end{equation}
where ${\cal D}_t^{1-\alpha}\to[s-ipU(x)]^{1-\alpha}$ in Laplace $s$ space. The derivation of Eq.
\eqref{forward_eq_fourier} is similar to that of \eqref{forward_eq_derive} (see \cite{BarkaiJSP10}
for more details).

\emph{Time-dependent force.---} Anomalous diffusion with a time-dependent force is of recent
interest
\cite{TimeDependentSokolov,TimeDependentHanggi,TimeDependentWeron,TimeDependentHenry,FriedrichEPL09}.
When the force is time-dependent, we assume the probabilities of jumping left and right are
determined by the force at the end of the waiting period
\cite{TimeDependentSokolov,TimeDependentHenry}. As we show in Appendix A, the equation for the PDF
$G(x,p,t)$ is similar to Eq. \eqref{forward_eq_derive}:
\begin{equation}
\label{forward_eq_timedep} \frac{\di}{\di t}G(x,p,t)=\Ka\Lfpt{\cal
D}_t^{1-\alpha}G(x,p,t)-pU(x)G(x,p,t),
\end{equation}
but where
\begin{equation*}
\Lfpt=\frac{\di^2}{\di x^2}-\frac{\di}{\di x}\frac{F(x,t)}{k_{B}T}
\end{equation*}
is the time-dependent Fokker-Planck operator. For $p=0$, Eq. \eqref{forward_eq_timedep} reduces to
the recently derived equation for the PDF of $x$ \cite{TimeDependentHenry}.

\subsection{The backward equation}

\label{sect_backward}

The forward equation describes $G(x,A,t)$, the joint PDF of $x$ and $A$. Consequently, if we are
interested only in the distribution of $A$, we must integrate $G$ over all $x$, which could be
inconvenient. We therefore develop below an equation for $G_{x_0}(A,t)$--- the PDF of $A$ at time
$t$, given that the process has started at $x_0$. This equation, which is called the backward
equation, turns out very useful in practical applications (see, e.g.,
\cite{BarkaiJSP10,MajumdarReview} and Section \ref{sect_occ_box}).

According to the CTRW model, the particle starts at $x=x_0$ and jumps at time $\tau$ to either
$x_0+a$ or $x_0-a$. Alternatively, the particle does not move at all during the measurement time
$[0,t]$. Hence,
\begin{align}
\label{backward_recursion} &G_{x_0}(A,t) = W(t)\delta[A-tU(x_0)] \\
&+\int_0^t
\psi(\tau)R(x_0)G_{x_0+a}[A-\tau U(x_0),t-\tau]d\tau \nonumber \\
&+\int_0^t \psi(\tau)L(x_0)G_{x_0-a}[A-\tau U(x_0),t-\tau]d\tau \nonumber.
\end{align}
Here, $\tau U(x_0)$ is the contribution to $A$ from the pausing time at $x_0$ in the time interval
$[0,\tau]$. The first term on the rhs of Eq. \eqref{backward_recursion} describes a motionless
particle, for which $A(t)=tU(x_0)$. We now transform Eq. \eqref{backward_recursion} $(x_0,A,t) \to
(k_0,p,s)$, using techniques similar to those used in Section \ref{sect_forward_derivation}. In the
continuum limit, $a\to 0$, this leads to,
\begin{align*}
G_{k_0}(p,s) &\approx \frac{1-\hat\psi\left[s+pU\left(-i\frac{\di}{\di
k_0}\right)\right]}{s+pU\left(-i\frac{\di}{\di k_0}\right)}\delta(k_0)\\
&+\hat\psi\left[s+pU\left(-i\frac{\di}{\di k_0}\right)\right]\times \nonumber
\\ &\left[\cos(k_0a)-\frac{aF\left(-i\frac{\di}{\di k_0}\right)}{2k_{B}T}i\sin(k_0a)\right]G_{k_0}(p,s)
\nonumber.
\end{align*}
We then expand $\hat{\psi}(s)\approx 1-B_{\alpha}s^{\alpha}$, $\cos(k_0a) \approx 1-k_0^2a^2/2$,
and $\sin(k_0a)\approx k_0a$. After some rearrangements,
\begin{align*}
&sG_{k_0}(p,s)-\delta(k_0) = -pU\left(-i\frac{\di}{\di
k_0}\right)G_{k_0}(p,s)\\&-\Ka\left[s+pU\left(-i\frac{\di}{\di k_0}\right)\right]^{1-\alpha}\times
\nonumber\\&\left[{k_0}^2+\frac{F\left(-i\frac{\di}{\di
k_0}\right)}{k_{B}T}ik_0\right]G_{k_0}(p,s)\nonumber.
\end{align*}
Inverting $k_0\to x_0$ and $s\to t$, we obtain the backward fractional Feynman-Kac equation:
\begin{equation}
\label{backward_eq} \frac{\di}{\di t}G_{x_0}(p,t)=\Ka{\cal
D}_t^{1-\alpha}{\Lfpb}G_{x_0}(p,t)-pU(x_0)G_{x_0}(p,t),
\end{equation}
where
\begin{equation}
\label{backward_fokker_planck_def} {\Lfpb}=\frac{\di^2}{\di
x_0^2}+\frac{F(x_0)}{k_{B}T}\frac{\di}{\di x_0}
\end{equation}
is the {\it backward} Fokker-Planck operator. The initial condition is $G_{x_0}(A,t=0)=\delta(A)$,
or $G_{x_0}(p,t=0)=1$. Note the $(+)$ sign of $\Lfpb$ and the order of the operators in its second
term, which are opposite to those of $\Lfp$ (Eq. \eqref{fokker-planck_def}). Here, ${\cal
D}_t^{1-\alpha}$ equals in Laplace $t\to s$ space $[s+pU(x_0)]^{1-\alpha}$. In Eq.
\eqref{forward_eq_derive} the operators depend on $x$ while in Eq. \eqref{backward_eq} they depend
on $x_0$. Therefore, Eq. \eqref{forward_eq_derive} is a forward equation while Eq.
\eqref{backward_eq} is a backward equation. Notice also that in Eq. \eqref{backward_eq}, the
fractional derivative operator appears to the left of the Fokker-Planck operator, in contrast to
the forward equation \eqref{forward_eq_derive}.

\section{The PDF of $\Ubar$ for long times}

For long measurement times, it is possible to use the fractional Feynman-Kac equation to obtain an
expression for the PDF of a general time-average:
\begin{equation*}
\Ubar=\frac{\int_0^t U[x(\tau)]d\tau}{t}=\frac{A}{t}.
\end{equation*}
We write first the forward equation \eqref{forward_eq_derive} in Laplace $s$ space:
\begin{align}
\label{forward_eq_ps} &\left[s+pU(x)\right]G(x,p,s)-G_0(x)\\
&=\Ka\left[\frac{\di^2}{\di x^2}-\frac{\di}{\di
x}\frac{F(x)}{k_{B}T}\right][s+pU(x)]^{1-\alpha}G(x,p,s) \nonumber.
\end{align}
CTRW functionals scale linearly with the time, $A\sim t$, and therefore, as shown in
\cite{GodrecheLuck}, $G(p,s)=g(p/s)/s$, where $g$ is a scaling function. Since we are interested in
the $t\to \infty$ limit, we take $s$ and $p$ to be small, with their ratio finite. We therefore
expect $G(x,p,s)\sim s^{-1}$ (indeed, see Eq. \eqref{G_statioanry_x} below), and consequently, both
terms on the lhs of \eqref{forward_eq_ps} scale as $s^0$. However, the rhs of \eqref{forward_eq_ps}
scales as $s^{-\al}$, and therefore for small $s$ the lhs is negligible. The forward equation thus
reduces to
\begin{equation*}
\label{eq_stationary_1} \Ka\left[\frac{\di^2}{\di x^2}-\frac{\di}{\di
x}\frac{F(x)}{k_{B}T}\right][s+pU(x)]^{1-\alpha}G(x,p,s) = 0.
\end{equation*}
The solution of the last equation is
\begin{equation}
\label{G_statioanry_const} G(x,p,s)= C(p,s)[s+pU(x)]^{\alpha-1}\exp\left[-\frac{V(x)}{k_BT}\right],
\end{equation}
where $C(p,s)$ is independent of $x$. To find $C$, we integrate Eq. \eqref{forward_eq_ps} over all
$x$:
\begin{equation}
\label{G_stationary_normalization} \int_{-\infty}^{\infty}\left[s+pU(x)\right]G(x,p,s)dx-1=0,
\end{equation}
which is true, because for a binding field, $G(x,p,s)$ and its derivative vanish for large $|x|$.
Substituting $G$ from Eq. \eqref{G_statioanry_const} into Eq. \eqref{G_stationary_normalization}
gives
\begin{equation*}
C(p,s)=\left\{\int_{-\infty}^{\infty}[s+pU(x)]^{\alpha}\exp\left[-\frac{V(x)}{k_BT}\right]dx\right\}^{-1}.
\end{equation*}
Therefore,
\begin{equation}
\label{G_statioanry_x}
G(x,p,s)=\frac{[s+pU(x)]^{\alpha-1}\exp\left[-\frac{V(x)}{k_BT}\right]}{\int_{-\infty}^{\infty}[s+pU(x)]^{\alpha}\exp\left[-\frac{V(x)}{k_BT}\right]dx}.
\end{equation}
Integrating Eq. \eqref{G_statioanry_x} over all $x$,
\begin{equation}
\label{G_stationary_ps}
G(p,s)=\frac{\int_{-\infty}^{\infty}[s+pU(x)]^{\alpha-1}\exp\left[-\frac{V(x)}{k_BT}\right]dx}{\int_{-\infty}^{\infty}[s+pU(x)]^{\alpha}\exp\left[-\frac{V(x)}{k_BT}\right]dx},
\end{equation}
where $G(p,s)$ is the double Laplace transform of $G(A,t)$, the PDF of $A$ at time $t$. The last
equation is the continuous version of the result derived using a different approach in
\cite{BarkaiPRL07,BarkaiJSP08}. As in \cite{BarkaiPRL07,BarkaiJSP08}, Eq. \eqref{G_stationary_ps}
can be inverted, using the method of \cite{GodrecheLuck}, to give the PDF of $\Ubar=A/t$,
\begin{align}
\label{G_stationary} G(\Ubar)&=\frac{\sin(\pi\al)}{\pi}\times\\&
\frac{I_{\al-1}^<(\Ubar)I_{\al}^>(\Ubar)+I_{\al-1}^>(\Ubar)I_{\al}^<(\Ubar)}{[I_{\al}^>(\Ubar)]^2+[I_{\al}^<(\Ubar)]^2+2\cos(\pi\al)I_{\al}^>(\Ubar)I_{\al}^<(\Ubar)}\nonumber,
\end{align}
where
\begin{equation*}
I_{\al}^{<}(\Ubar)=\int_{\Ubar<U(x)}\exp\left[-\frac{V(x)}{k_BT}\right]\left[U(x)-\Ubar\right]^{\al}dx
\end{equation*}
and
\begin{equation*}
I_{\al}^{>}(\Ubar)=\int_{\Ubar>U(x)}\exp\left[-\frac{V(x)}{k_BT}\right]\left[\Ubar-U(x)\right]^{\al}dx.
\end{equation*}
For normal diffusion, $\al=1$, the PDF is a delta function
$G(\Ubar)=\delta\left[\Ubar-\av{U}_{\textrm{th}}\right]$ \cite{BarkaiPRL07,BarkaiJSP08}. For
anomalous sub-diffusion, $\al<1$, $\Ubar$ is a random variable, different from the ensemble
average. This behavior of the time-average results from the weak ergodicity breaking of the
sub-diffusing system. Similar results hold when $U(x)$ is not necessarily positive: the Laplace
transform $A\to p$ is replaced by a Fourier transform and in Eq. \eqref{G_stationary_ps}, $p$ is
replaced by $-ip$.

\section{Applications: Weak ergodicity breaking}

In this section we present two applications of the fractional Feynman-Kac equation: the occupation
fraction in a box and the time-averaged position in a harmonic potential. We demonstrate weak
ergodicity breaking in both cases and investigate the convergence to the asymptotic limits.

\subsection{The occupation fraction in the positive half of a box}

\label{sect_occ_box}

We study the problem of the occupation time in $x>0$ for a sub-diffusing particle moving freely in
the box extending between $\left[-\frac{L}{2},\frac{L}{2}\right]$
\cite{BarkaiPRL05,BarkaiJSP06,BarkaiJSP08}.

\subsubsection{The distribution}

\label{occ_time_dist}

Define the occupation time in $x>0$ as $T_+=\int_0^t\Theta[x(\tau)]d\tau$ (namely
$U(x)=\Theta(x)$). To find the PDF of $T_+$, we write the backward fractional Feynman-Kac equation
\eqref{backward_eq} in Laplace $s$ space:
\begin{align}
\label{backward_eq_occupation} &sG_{x_0}(p,s)-1=\\&\begin{cases}\Ka s^{1-\alpha}\dixzs
G_{x_0}(p,s)&x_0<0,\\\Ka (s+p)^{1-\alpha}\dixzs G_{x_0}(p,s)-pG_{x_0}(p,s)&x_0>0.\end{cases}
\nonumber
\end{align}
The equation \eqref{backward_eq_occupation} is subject to the boundary conditions:
\begin{equation*}
\left.\dixz G_{x_0}(p,s)\right|_{x_0=\pm\Lh}=0.
\end{equation*}
The solution of the last equation is:
\begin{equation}
\label{backward_eq_occ_sol} G_{x_0}(p,s)=\begin{cases}
C_0\cosh\left[\left(\frac{L}{2}+x_0\right)\frac{s^{\alpha/2}}{\sKa}\right]+\frac{1}{s} & x_0<0, \\
C_1\cosh\left[\left(\frac{L}{2}-x_0\right)\frac{(s+p)^{\alpha/2}}{\sKa}\right]+\frac{1}{s+p} &
x_0>0.
\end{cases}
\end{equation}
Matching $G$ and its derivative at $x_0=0$ gives the equations:
\begin{align*}
&C_0\cosh\left(\frac{Ls^{\alpha/2}}{2\sKa}\right)+\frac{1}{s}=C_1\cosh\left[\frac{L(s+p)^{\alpha/2}}{2\sKa}\right]+\frac{1}{s+p},\nonumber \\
&C_0s^{\alpha/2}\sinh\left(\frac{Ls^{\alpha/2}}{2\sKa}\right)=-C_1(s+p)^{\alpha/2}\sinh\left[\frac{L(s+p)^{\alpha/2}}{2\sKa}\right].
\end{align*}
Solving these equations for $C_0$ and $C_1$ and substituting $x_0=0$ in Eq.
\eqref{backward_eq_occ_sol} gives, after some algebra,
\begin{align}
\label{sol_occ_box_ps_allt} &G_0(p,s)=\\&\frac{s^{\alpha/2-1}\tanh\left[(s\tau)^{\alpha/2}\right]+
(s+p)^{\alpha/2-1}\tanh\left\{\left[\tau(s+p)\right]^{\alpha/2}\right\}}
{s^{\alpha/2}\tanh\left[(s\tau)^{\alpha/2}\right]+
(s+p)^{\alpha/2}\tanh\left\{\left[\tau(s+p)\right]^{\alpha/2}\right\}},\nonumber
\end{align}
where we defined $\tau^{\al}\equiv L^2/(4\Ka)$. This equation was previously derived in
\cite{BarkaiJSP06} using a different method. Eq. \eqref{sol_occ_box_ps_allt} describes the PDF of
$T_+$ for all times, but cannot be directly inverted. For long times, or $(s\tau)^{\al/2}\ll 1$,
\begin{equation}
\label{Lamperti_ps} G_0(p,s)\approx\frac{s^{\alpha-1}+ (s+p)^{\alpha-1}} {s^{\alpha}+
(s+p)^{\alpha}}.
\end{equation}
This can be inverted to give the PDF of $\lambda\equiv T_+/t$, or the occupation fraction
\cite{GodrecheLuck,BarkaiJSP06},
\begin{equation}
\label{Lamperti_eq}
G(\lambda)=\frac{\sin(\pi\alpha)}{\pi}\frac{\lambda^{\alpha-1}(1-\lambda)^{\alpha-1}}{\lambda^{2\alpha}+(1-\lambda)^{2\alpha}+2\cos(\pi\alpha)\lambda^{\alpha}(1-\lambda)^{\alpha}}.
\end{equation}
Eq. \eqref{Lamperti_eq} is called Lamperti's PDF \cite{Lamperti}. Note that Eqs.
\eqref{Lamperti_ps} and \eqref{Lamperti_eq} can also be derived directly from the general
long-times limit, Eqs. \eqref{G_stationary_ps} and \eqref{G_stationary}, respectively. Whereas the
PDF of the occupation fraction for a free particle is also Lamperti's
\cite{BarkaiJSP06,BarkaiJSP10}, in the free particle case the exponent is $\alpha/2$, compared to
$\alpha$ here. An equation for $G_{x_0}(p,s)$ for $x_0\ne 0$ can be derived in exactly the same
manner, leading, for long times, to Eqs. \eqref{Lamperti_ps} and \eqref{Lamperti_eq}, as expected.

For $\alpha=1$, it is easy to see from Eq. \eqref{Lamperti_ps} that $G(T_+,t)=\delta(T_+-t/2)$ or
$\lambda=1/2$. This is the expected result based on the ergodicity of normal diffusion. As $\al$
decreases below 1, the delta function spreads out to form a W shape. For even smaller values of
$\alpha$ ($\lesssim 0.59$ \cite{MargolinJSP}), the peak at $\lambda=1/2$ disappears and the PDF
attains a U shape, indicating that the particle spends almost its entire time in only one of the
half-boxes. For $\alpha\to 0$, $G(\lambda)=\delta(\lambda)/2+\delta(\lambda-1)/2$, as expected.
This behavior is demonstrated and compared to simulations in Figure \ref{Fig_occ_PDF}. Details on
the simulation method are given in Appendix B.

For short times, $(t/\tau)^{\alpha/2}\ll 1$, we substitute in Eq. \eqref{sol_occ_box_ps_allt} the
limit $(s\tau)^{\alpha/2}\gg 1$,
\begin{equation}
\label{Lamperti_ps_short} G_0(p,s)\approx\frac{s^{\alpha/2-1}+ (s+p)^{\alpha/2-1}} {s^{\alpha/2}+
(s+p)^{\alpha/2}}.
\end{equation}
In $t$ space, this gives again the Lamperti PDF, but now with index $\alpha/2$. This is exactly the
PDF of the occupation fraction of a free particle, which is expected, because for short times the
particle does not interact with the boundaries \cite{BarkaiJSP06}. It can be shown that for short
times, $G_{x_0>0}(T_+,t)=\delta(T_+-t)$, and $G_{x_0<0}(T_+,t)=\delta(T_+)$, as expected.

\begin{figure}
\begin{center}
\epsfig{file=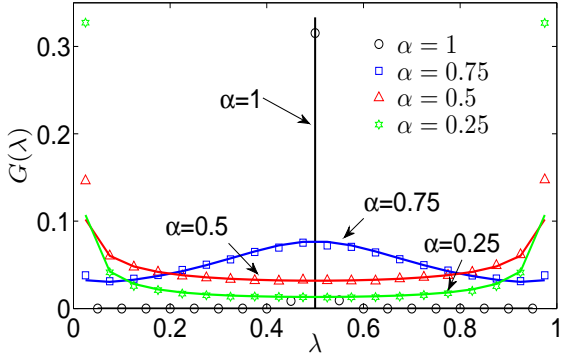,height=5cm,width=8cm}
\end{center} \caption{The PDF of the occupation fraction in half-space for a particle in the box $\left[-1,1\right]$.
CTRW trajectories were generated as explained in Appendix B, with $x_0=0$. For each trajectory, the
total time in $x>0$, $T_+$, was recorded, and the occupation fraction, $\lambda=T_+/t$, was
calculated. The figure shows the PDF for long times of the occupation fraction $\lambda$ for
$\alpha=1,0.75,0.5,0.25$ (symbols). Lamperti's PDF, Eq. \eqref{Lamperti_eq}, is plotted as lines
(For $\al=1$, the PDF for the simulations and the theory was scaled by 3 for visibility). While for
$\alpha=1$, $\lambda$ is very narrowly distributed around $1/2$, for $\alpha<1$, the PDF becomes
wider and even attains a U shape for small enough $\alpha$.} \label{Fig_occ_PDF}
\end{figure}

\subsubsection{An application of the occupation time functional--- the first passage time PDF}

As a side note, we demonstrate how the fractional Feynman-Kac equation for the occupation time can
be applied in an elegant manner to the problem of the first passage time (FPT). The FPT in the box
$\left[-\frac{L}{2},\frac{L}{2}\right]$ is defined as the time $t_f$ it takes a particle starting
at $x_0=-b\;(0<b<L/2)$ to reach $x=0$ for the first time \cite{Redner_book}. A relation between the
occupation time functional of the previous subsection and the FPT was proposed by Kac
\cite{Kac1951}:
\begin{equation*}
\textrm{Pr}\{t_f>t\}=\textrm{Pr}\{\max_{0\leq \tau \leq t}x(\tau)<0\} = \lim_{p\to
\infty}G_{x_0}(p,t),
\end{equation*}
where as in the previous subsection, $G_{x_0}(p,s)$ is the PDF of
$T_+=\int_0^t\Theta[x(\tau)]d\tau$. The last equation is true since
$G_{x_0}(p,t)=\int_0^{\infty}e^{-pT_+}G_{x_0}(T_+,t)dT_+$, and thus, if the particle has never
crossed $x=0$, we have $T_+=0$ and $e^{-pT_+}=1$, while otherwise, $T_+>0$ and for $p\to \infty$,
$e^{-pT_+}=0$. Substituting $x_0=-b$ and $p\to \infty$ in Eq. \eqref{backward_eq_occ_sol} of the
previous subsection gives
\begin{equation}
\label{Survival_box_ps} \lim_{p\to
\infty}G_{-b}(p,s)=\frac{1}{s}\left\{1-\frac{\cosh\left[\left(\frac{L}{2}-b\right)\frac{s^{\alpha/2}}{\sKa}\right]}{\cosh\left(\frac{Ls^{\alpha/2}}{2\sKa}\right)}\right\}.
\end{equation}
The first passage time PDF satisfies $f(t)=\frac{\di}{\di t}\left[1-\textrm{Pr}\{t_f>t\}\right]$.
We therefore have in Laplace space,
\begin{equation*}
f(s)=\frac{\cosh\left[\left(\frac{L}{2}-b\right)\frac{s^{\alpha/2}}{\sKa}\right]}{\cosh\left(\frac{Ls^{\alpha/2}}{2\sKa}\right)}.
\end{equation*}
For long times, the small $s$ limit gives
\begin{equation*}
f(s)\approx 1-\frac{b(L-b)}{2\Ka}s^{\alpha}.
\end{equation*}
For $0<\al<1$, inverting $s\to t$,
\begin{equation}
\label{FPT_box} f(t_f)\approx\frac{b(L-b)}{2\Ka|\Gamma(-\alpha)|}t_f^{-(1+\alpha)}.
\end{equation}
Therefore, $f(t_f)\sim t_f^{-(1+\alpha)}$ (compared to $f(t_f)\sim t_f^{-(1+\alpha/2)}$ for a free
particle \cite{BarkaiPRE01,BarkaiJSP10}), indicating that for $\alpha<1$, $\av{t_f}=\infty$.
Eqs. \eqref{Survival_box_ps} and \eqref{FPT_box} agree with previous work
\cite{BarkaiJSP06,BarkaiPRE06}.

\subsubsection{The fluctuations}

Eq. \eqref{sol_occ_box_ps_allt}, giving $G_0(p,s)$ for the occupation time functional, cannot be
directly inverted. It can nevertheless be used to calculate the first few moments using
\begin{equation*}
\av{T_+^n}=(-1)^n\left.\frac{\di^n}{\di p^n}G_0(p,t)\right|_{p=0}.
\end{equation*}
The first moment (for $x_0=0$) is of course $\av{T_+}=t/2$ or $\av{\lambda}=1/2$. For the second
moment,
\begin{equation}
\label{Fluc_occ_all_s} \av{T_+^2}_s=\frac{4-\alpha}{4s^3}-\frac{\alpha (s\tau)^{\alpha/2}}{2
s^3\sinh\left[2(s\tau)^{\alpha/2}\right]}.
\end{equation}
The long times, we take the limit of small $s$,
\begin{equation*}
\av{T_+^2}_s\approx\frac{2-\alpha}{2s^3}+\frac{\alpha \tau^{\alpha}}{6s^{3-\alpha}}.
\end{equation*}
Inverting and dividing by $t^2$, we obtain the fluctuations of the occupation fraction,
$\av{(\Delta\lambda)^2}=\av{\lambda^2}-\av{\lambda}^2$,
\begin{equation}
\label{Fluc_occ_frac}
\av{(\Delta\lambda)^2}\approx\frac{1-\alpha}{4}+\frac{\alpha}{6\Gamma(3-\alpha)}\left(\frac{t}{\tau}\right)^{-\alpha}.
\end{equation}
For $\alpha<1$ and $t\to \infty$, we see from Eq. \eqref{Fluc_occ_frac} that
$\av{(\Delta\lambda)^2}=\frac{1-\alpha}{4}>0$. For $\alpha=1$, $\av{(\Delta\lambda)^2}\to 0$ as
$t\to \infty$. The convergence to the long-times limit exhibits a $t^{-\alpha}$ decay. For $x_0\ne
0$, the first moment approaches $1/2$ as $\av{\lambda}\approx
1/2+\frac{x_0(L-|x_0|)}{4\Ka\Gamma(2-\al)}t^{-\alpha}$ and the fluctuations remain the same as in
Eq. \eqref{Fluc_occ_frac} up to order $t^{-\alpha}$.

For short times (and $x_0=0$), taking the limit $(s\tau)^{\al/2}\gg 1$ in Eq.
\eqref{Fluc_occ_all_s} gives $\av{T_+^2}_s\approx\frac{4-\alpha}{4s^3}$, from which
\begin{equation}
\label{Fluc_occ_frac_short} \av{(\Delta\lambda)^2} \approx \frac{1-\alpha/2}{4}.
\end{equation}
This is the expected result, since for short times the PDF is Lamperti's with index $\alpha/2$ (Eq.
\eqref{Lamperti_ps_short}).

\begin{figure}
\begin{center}
\epsfig{file=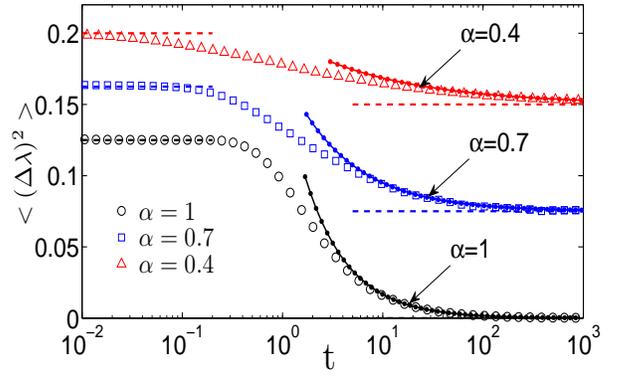,height=5.25cm,width=8cm}
\end{center} \caption{The fluctuations of the occupation fraction in half box.
CTRW trajectories were generated as explained in Appendix B (with $x_0=0$) and the occupation
fraction in half box, $\lambda=T_+/t$, was calculated. The figure shows the fluctuations
$\av{(\Delta\lambda)^2}$ vs. $t$ for $\alpha=0.4,0.7,1$ (symbols). Theory for long times, Eq.
\eqref{Fluc_occ_frac}, is plotted as dotted lines. The fluctuations are initially equal to their
free particle counterpart, $(1-\alpha/2)/4$ (Eq. \eqref{Fluc_occ_frac_short}, indicated as dashed
lines), and then decay to their asymptotic value, $(1-\alpha)/4$ (also indicated as dashed lines),
as $t^{-\alpha}$. Only for $\alpha=1$, the fluctuations vanish for $t\to\infty$.}
\label{Fig_occ_moments}
\end{figure}

The fluctuations $\av{(\Delta\lambda)^2}$ are plotted in Figure \ref{Fig_occ_moments} and agree
well with Eq. \eqref{Fluc_occ_frac_short} for short times and with Eq. \eqref{Fluc_occ_frac} for
long times. As expected, the approach to the asymptotic limit is slower as $\al$ becomes smaller.

\subsection{The time-averaged position in a harmonic potential}

We consider the time-averaged position, $\xbar=\frac{1}{t}\int_0^t x(\tau)d\tau$, for a
sub-diffusing particle in a harmonic potential, $V(x)=m\omega^2x^2/2$ (fractional
Ornstein-Uhlenbeck process \cite{BarkaiPRL99,BarkaiJPC00}).

\subsubsection{The distribution}

\begin{figure}
\begin{center}
\epsfig{file=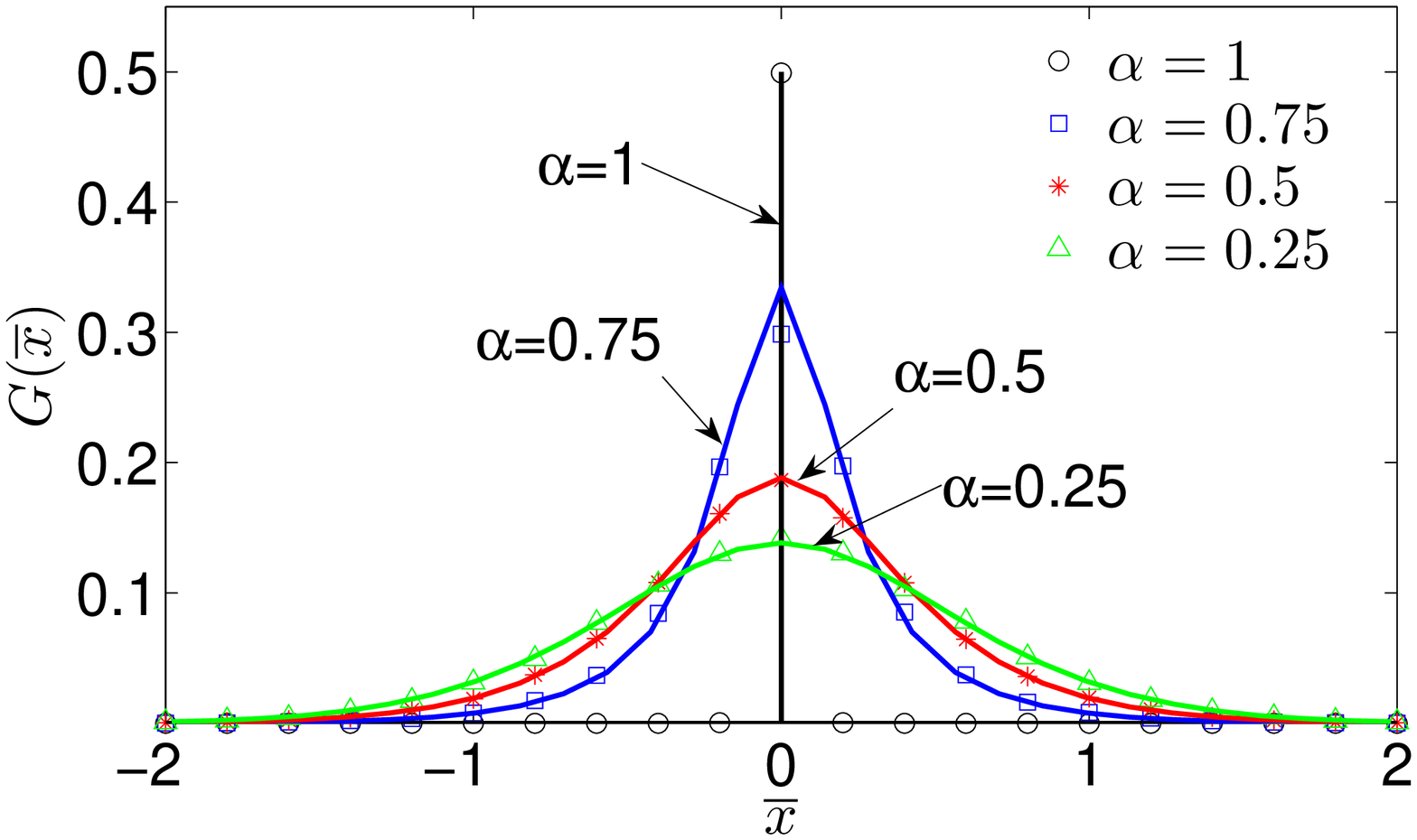,height=4.75cm,width=8cm}
\epsfig{file=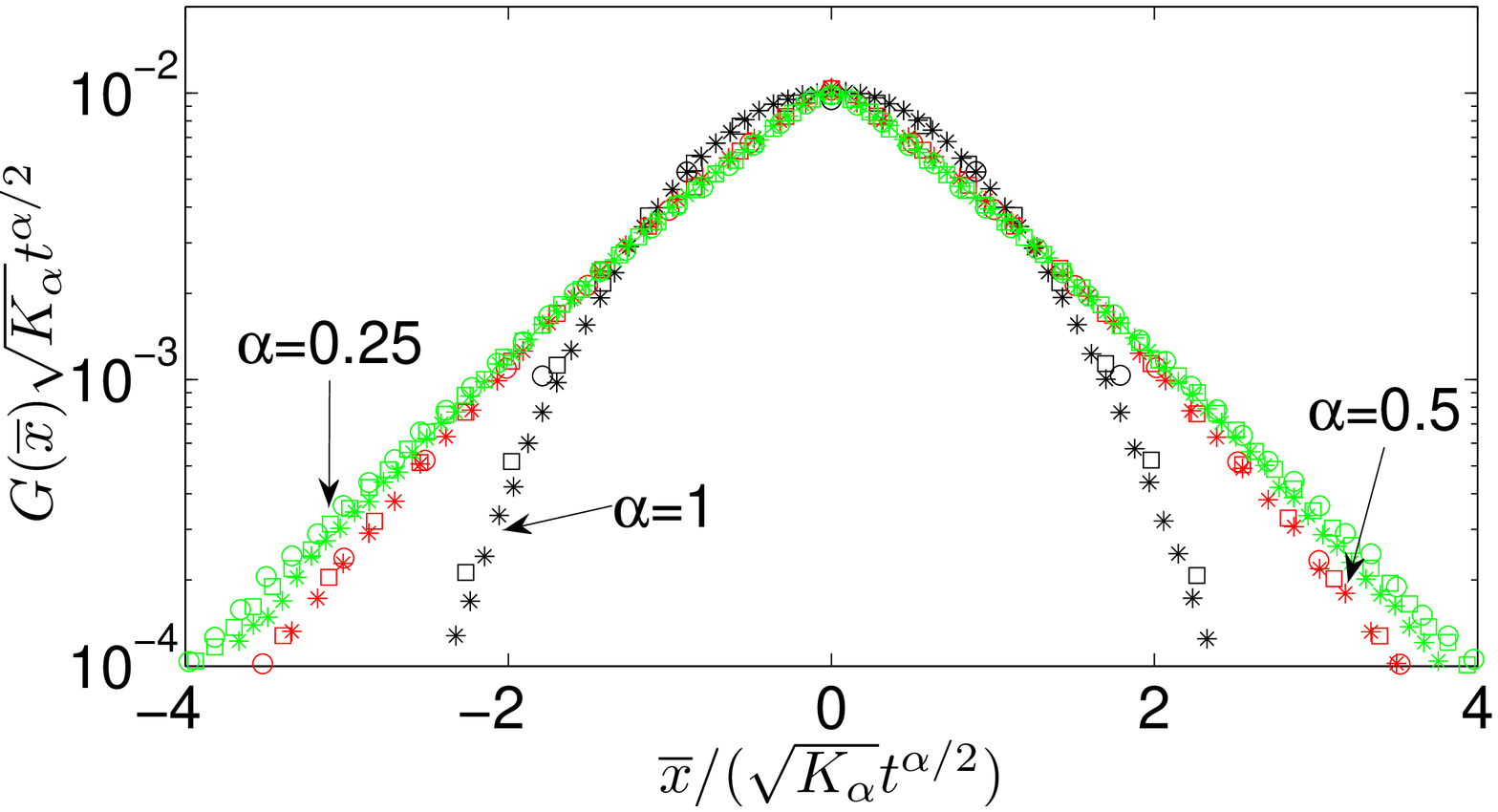,height=4.75cm,width=8cm}
\end{center}
\caption{The PDF $G(\xbar)$ for a particle in a binding harmonic field. CTRW trajectories were
generated using the method described in Appendix B, with $x_0=0$. Top panel: Simulation results for
long times for $\alpha=0.25,0.5,0.75,1$ (symbols). Theory for $t\to \infty$, Eq.
\eqref{G_stationary_xbar}, is plotted as solid lines (For $\al=1$, the PDF for the simulations and
the theory was scaled by 2 for visibility). For $\alpha=1$, the distribution is a delta function,
whereas for $\alpha<1$, $\xbar$ is a random variable even for long times, indicating ergodicity
breaking. Bottom panel: Simulation results for the PDF of $\xbar$ for a number of short times and
for $\al=0.25,0.5,1$ (symbols). The plot illustrates the free-particle scaling form, Eq.
\eqref{xbar_pdf_short}.} \label{Fig_xbar_PDF}
\end{figure}

We first study the PDF in the long-times limit using the general equation \eqref{G_stationary}.
Define the second moment in thermal equilibrium as $\x2th=k_BT/(m\omega^2)$. Measuring $\xbar$ in
units of $\sqrt{\x2th}$, we have for $t\to \infty$,
\begin{equation*}
G(\xbar)=\frac{1}{\sqrt{\x2th}}g\left(\frac{\xbar}{\sqrt{\x2th}}\right),
\end{equation*}
where
\begin{equation}
\label{G_stationary_xbar}
g(y)=\frac{\sin(\pi\al)}{\pi}\frac{I_{\al-1}^<(y)I_{\al}^>(y)+I_{\al-1}^>(y)I_{\al}^<(y)}{[I_{\al}^>(y)]^2+[I_{\al}^<(y)]^2+2\cos(\pi\al)I_{\al}^>(y)I_{\al}^<(y)},
\end{equation}
with
\begin{equation*}
I_{\al}^<=\int_{y}^{\infty}e^{-\frac{x^2}{2}}(x-y)^{\al}dx\;;\;I_{\al}^>=\int_{-\infty}^{y}e^{-\frac{x^2}{2}}(y-x)^{\al}dx.
\end{equation*}
Using Mathematica, we could express the solution of the integrals in Eq. \eqref{G_stationary_xbar}
in terms of Kummer's functions. The full expression is given in Appendix C (Eq.
\eqref{G_stationary_xbar_full}). It can be shown that for $\alpha=1$, $G(\xbar)=\delta(\xbar)$, as
expected for an ergodic system \cite{BarkaiPRL07,BarkaiJSP08}. For $\al<1$, $G(\xbar)$ has a
non-zero width, and when $\alpha\to 0$, $G(\xbar)=\sqrt{\frac{m\omega^2}{2\pi
k_BT}}\exp\left[-\frac{m\omega^2\xbar^2}{2k_BT}\right]$, which is the Boltzmann distribution, since
for $\alpha\to 0$, $\xbar \to x$ \cite{BarkaiPRL07,BarkaiJSP08}. For $\xbar\ll\sqrt{\x2th}$ ($y \ll
1$), $g(y)$ has a Taylor expansion around $y=0$ of the form
$g(y)=\frac{\Gamma\left(\frac{\alpha}{2}\right)\tan\left(\frac{\pi\al}{2}\right)}{\sqrt{2}\pi\Gamma\left(\frac{1+\al}{2}\right)}+{\cal
O}(y^2)$. For $\xbar\gg\sqrt{\x2th}$ ($y \gg 1$), $g(y)\sim
\frac{\Gamma(\al)\sin(\pi\al)}{\sqrt{2\pi^3}}y^{-2\al}e^{-y^2/2}$, which gives the expected results
for $\al\to 0$ and $\al=1$. Eq. \eqref{G_stationary_xbar} is plotted and compared to simulations in
Fig \ref{Fig_xbar_PDF}.

For \emph{short} times, $t^{\al}\ll \x2th/\Ka$,
the particle is at the minimum of the potential and therefore
behaves as a free particle. For the free-particle case, we have previously shown a scaling form for
$x_0=0$ \cite{BarkaiJSP10}
\begin{equation}
\label{xbar_pdf_short} G(\xbar,t)=\frac{1}{\sKa t^{\alpha/2}}h_{\alpha}\left(\frac{\xbar}{\sKa
t^{\alpha/2}}\right),
\end{equation}
where $h_{\alpha}(y)$ is a dimensionless scaling function. This behavior is numerically
demonstrated in Fig \ref{Fig_xbar_PDF}.

\subsubsection{The fluctuations}

The PDF of the time-averaged position was shown in the previous subsection to have a non-trivial
limiting distributions for $t\to \infty$ (Eq. \eqref{G_stationary_xbar}) and $t\to 0$ (Eq.
\eqref{xbar_pdf_short}). However, the shape of the PDF for other times is unknown. In this
subsection, we show that using the fractional Feynman-Kac equation, we can determine the width of
the distribution for all times.

\begin{figure}
\begin{center}
\epsfig{file=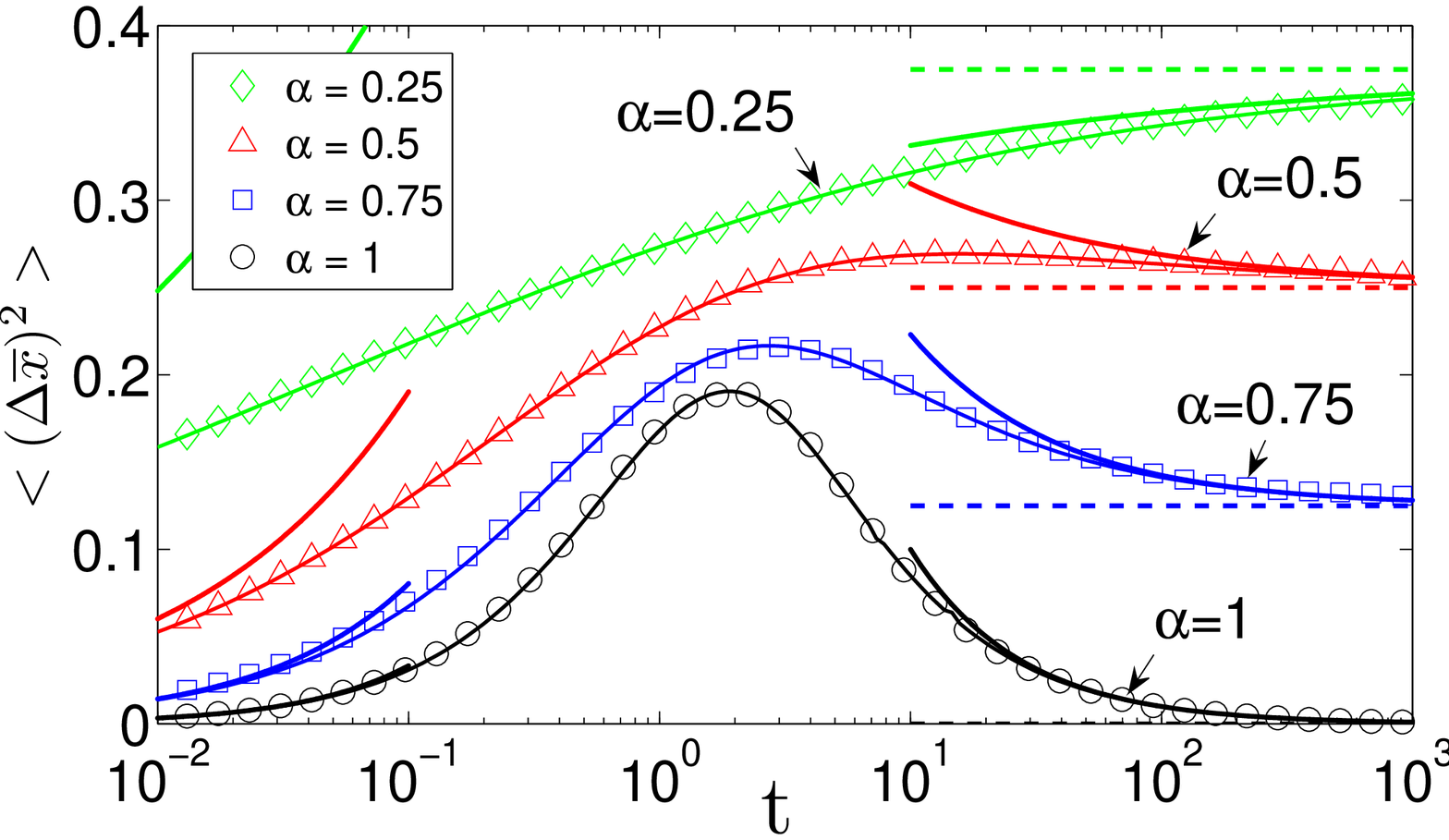,height=4.8cm,width=8cm}
\epsfig{file=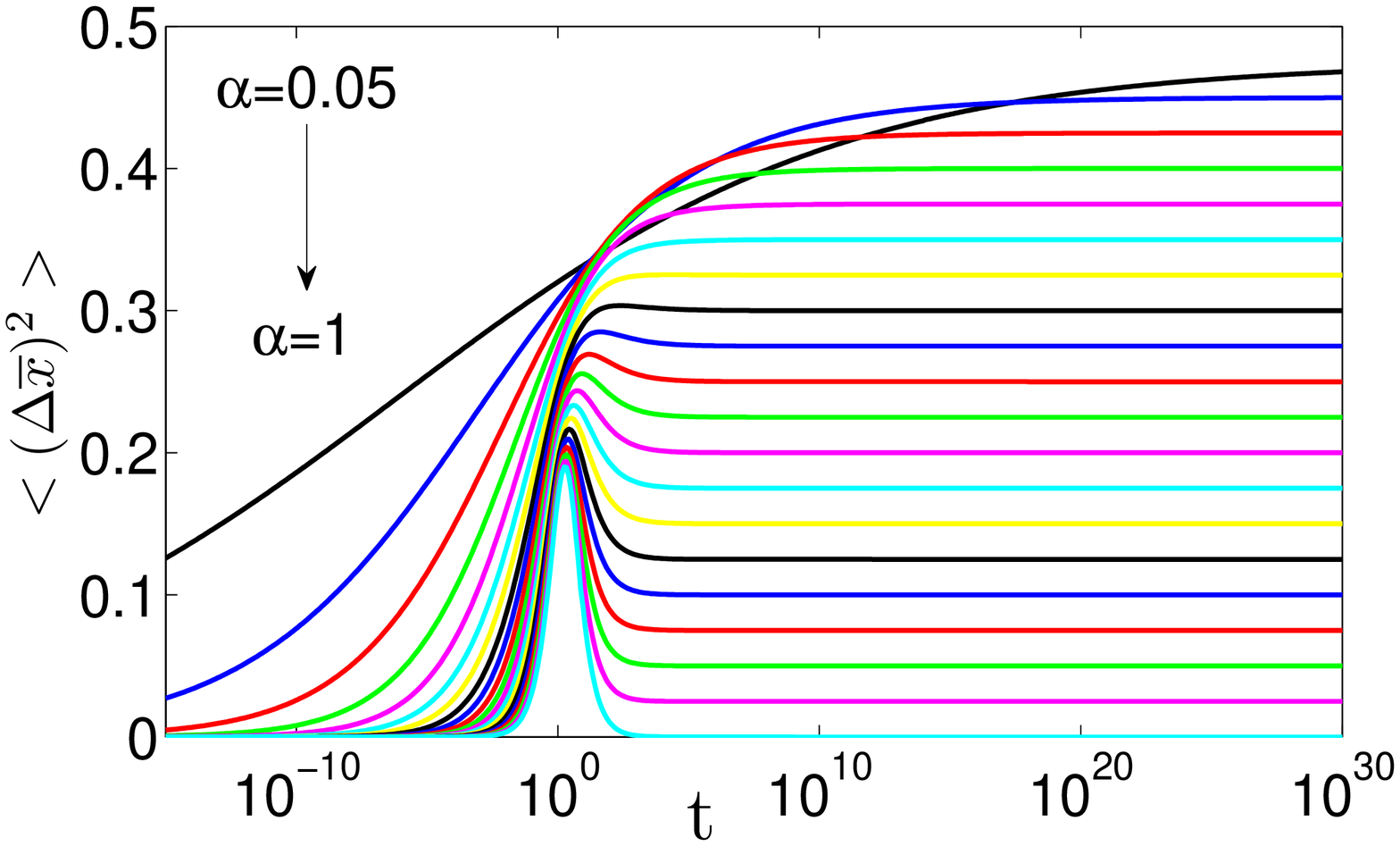,height=4.8cm,width=8cm}
\end{center}
\caption{The fluctuations $\av{(\Delta\xbar)^2}$ for a particle in a harmonic potential. Top panel:
CTRW trajectories were generated using the method described in Appendix B, with $x_0=0$. Symbols
represent simulation results for $\alpha=0.25,0.5,0.75,1$. Theory, Eq. \eqref{timeavg_fluc_mittag},
is plotted as solid lines. The straight dashed lines are $\lim_{t\to
\infty}\av{(\Delta\xbar)^2}=(1-\alpha)\x2th$. Except for $\alpha=1$, the fluctuations do not vanish
when $t\to \infty$ and thus ergodicity is broken. The dotted lines represent the long-times and
short-times approximations, Eqs. \eqref{xbar_avg_longt} and \eqref{xbar_avg_shortt}, respectively.
Bottom panel: The fluctuations, Eq. \eqref{timeavg_fluc_mittag}, plotted for a wide time range
$[10^{-15},10^{30}]$. Shown are 20 curves for $\al=0.05,0.1,0.15,...,1$ (top to bottom). The
fluctuations display a maximum when $\al<1/3$ and a crossover when $\al\lesssim 0.15$. As expected,
the fluctuations approach their asymptotic value slower for smaller values of $\al$.}
\label{Fig_xbar_fluc}
\end{figure}

Let us write the forward equation in $(p,s)$ space for the functional $A=\xbar t=\int_0^t
x(\tau)d\tau$ ($U(x)=x$) and for $x_0=0$. Since $A$ is not necessarily positive, $p$ here is the
Fourier pair of $A$ and we use Eq. \eqref{forward_eq_fourier} of Section \ref{sect_specialcases}:
\begin{align}
\label{eq_time_average_xps} &sG(x,p,s)-\delta(x)=ipxG(x,p,s)\\ &+\Ka\left[\frac{\di^2}{\di
x^2}+\frac{\di}{\di x}\frac{m\omega^2x}{k_{B}T}\right][s-ipx]^{1-\alpha}G(x,p,s)\nonumber.
\end{align}
To find $\av{A^2}$, we use the relation
\begin{equation*}
\av{A^2}_s=-\int_{-\infty}^{\infty}\left.\frac{\di^2}{\di p^2}G(x,p,s)\right|_{p=0}dx.
\end{equation*}
Operating on both sides of Eq. \eqref{eq_time_average_xps} with $-\frac{\di^2}{\di p^2}$,
substituting $p=0$, and integrating over all $x$, we obtain, in $s$ space,
\begin{align}
\label{eq_time_average_A_x2} s\av{A^2}_s=2\av{Ax}_s,
\end{align}
where we used the fact that the integral over the Fokker-Planck operator vanishes. Eq.
\eqref{eq_time_average_A_x2} can be intuitively understood by noting that $\frac{\di}{\di
t}\av{A^2}=2\av{A\dot{A}}$ and $\dot{A}=x$. We next use Eq. \eqref{eq_time_average_xps} and
\begin{equation*}
\av{Ax}_s=-i\int_{-\infty}^{\infty}x\left.\frac{\di}{\di p}G(x,p,s)\right|_{p=0}dx,
\end{equation*}
to obtain,
\begin{equation*}
s\av{Ax}_s=\left[1+(1-\alpha)(s\tau)^{-\alpha}\right]\av{x^2}_s-s(s\tau)^{-\alpha}\av{Ax}_s,
\end{equation*}
where we defined the relaxation time $\tau^{\alpha}=k_{B}T/(\Ka m\omega^2)=\x2th/\Ka$. Thus,
\begin{equation}
\label{eq_time_average_A_xx}
s\av{A_xx}_s=\frac{(1-\alpha)+(s\tau)^{\alpha}}{1+(s\tau)^{\alpha}}\av{x^2}_s.
\end{equation}
Finally, to find $\av{x^2}_s$, we use $\av{x^2}_s=\int_{-\infty}^{\infty}x^2G(x,p=0,s)dx$,
\begin{equation*}
s\av{x^2}_s=2\Ka s^{-\alpha}-2s(s\tau)^{-\alpha}\av{x^2}_s,
\end{equation*}
where we used the normalization condition $\int G(x,p=0,s)dx=1/s$. Thus,
\begin{equation}
\label{eq_time_average_x2} s\av{x^2}_s=\frac{2\x2th}{2+(s\tau)^{\alpha}}.
\end{equation}
Combining Eqs. \eqref{eq_time_average_A_x2}, \eqref{eq_time_average_A_xx}, and
\eqref{eq_time_average_x2}, we find,
\begin{equation*}
\av{A^2}_s=\frac{4}{s^3}\frac{(1-\alpha)+(s\tau)^{\alpha}}{1+(s\tau)^{\alpha}}
\frac{\x2th}{2+(s\tau)^{\alpha}}.
\end{equation*}
To invert to the time domain, we write $\av{A^2}_s$ as partial fractions:
\begin{align}
\label{time_average_partial_fractions} &\av{A^2}_s=\frac{2\x2th}{s^3}\times
\\&\left[(1-\alpha)+2\alpha\frac{(s\tau)^{\alpha}}{1+(s\tau)^{\alpha}}
-(1+\alpha)\frac{(s\tau)^{\alpha}}{2+(s\tau)^{\alpha}}\right] \nonumber.
\end{align}
Inverting the last equation, we find
\begin{align}
&\av{A^2}=\x2th t^2\times \\&\left\{(1-\alpha)+4\alpha
E_{\alpha,3}\left[-(t/\tau)^{\alpha}\right]-2(1+\alpha)E_{\alpha,3}\left[-2(t/\tau)^{\alpha}\right]\right\}\nonumber,
\end{align}
where we used the Laplace transform relation \cite{Podlubny}
\begin{equation*}
\int_0^{\infty}e^{-st}t^2E_{\alpha,3}\left[-c(t/\tau)^{\alpha}\right]dt=\frac{1}{s^3}\frac{(s\tau)^{\alpha}}{c+(s\tau)^{\alpha}},
\end{equation*}
and $E_{\alpha,3}(z)$ is the Mittag-Leffler function, defined as \cite{Podlubny}
\begin{equation*}
E_{\alpha,3}(z)=\sum_{n=0}^{\infty}\frac{z^n}{\Gamma(3+\alpha n)}.
\end{equation*}
To obtain the fluctuations of the time-averaged position,
$\av{(\Delta\xbar)^2}=\av{\xbar^2}-\av{\xbar}^2$, we use $\av{\xbar^2}=\av{A^2}/t^2$ and
$\av{\xbar}=0$ (since $x_0=0$). This gives
\begin{align}
\label{timeavg_fluc_mittag} &\av{(\Delta\xbar)^2}=\x2th\times \\&\left\{(1-\alpha)+4\alpha
E_{\alpha,3}\left[-(t/\tau)^{\alpha}\right]-2(1+\alpha)E_{\alpha,3}\left[-2(t/\tau)^{\alpha}\right]\right\}\nonumber.
\end{align}
Eq. \eqref{timeavg_fluc_mittag} is plotted (using \cite{PodlubnyMLF}) and compared to simulations
in the top panel of Figure \ref{Fig_xbar_fluc}.

To find the long times behavior of the fluctuations \eqref{timeavg_fluc_mittag}, we expand Eq.
\eqref{time_average_partial_fractions} for small $s$, invert, and divide by $t^2$,
\begin{equation}
\label{xbar_avg_longt} \av{(\Delta\overline{x})^2}\approx (1-\alpha)\x2th+\frac{(3\alpha-1)\x2th}
{\Gamma(3-\alpha)}\left(\frac{t}{\tau}\right)^{-\alpha}.
\end{equation}
Thus, for $\alpha<1$ and $t\to \infty$, $\av{(\Delta\overline{x})^2}=(1-\alpha)\x2th>0$ and
ergodicity is broken. Only when $\alpha=1$, we have ergodic behavior
$\av{(\Delta\overline{x})^2}=0$. As we observed for the occupation fraction (Eq.
\eqref{Fluc_occ_frac}), Eq. \eqref{xbar_avg_longt} too exhibits a $t^{-\alpha}$ convergence of the
fluctuations to their asymptotic limit.

For short times,
\begin{equation*}
\label{mlf_short} E_{\alpha,3}\left[-(t/\tau)^{\alpha}\right]\approx
\frac{1}{2}-\frac{(t/\tau)^{\alpha}}{\Gamma(3+\alpha)}.
\end{equation*}
Therefore,
\begin{equation}
\label{xbar_avg_shortt} \av{(\Delta\overline{x})^2}\approx
\frac{4\x2th}{\Gamma(3+\alpha)}\left(\frac{t}{\tau}\right)^{\alpha}.
\end{equation}
Noting that $\x2th/\tau^{\al}=\Ka$, we can rewrite Eq. \eqref{xbar_avg_shortt}, as
$\av{(\Delta\overline{x})^2}\approx \frac{4\Ka}{\Gamma(3+\alpha)}t^{\alpha}$, which is, as
expected, equal to the free particle expression \cite{BarkaiJSP10}.

The bottom panel of Figure \ref{Fig_xbar_fluc} presents the fluctuations of the time-average (for
$x_0=0$) for a wide range of times and for $\alpha=0.05,0.1,0.15,...,1$. As expected from Eqs.
\eqref{xbar_avg_longt} and \eqref{xbar_avg_shortt}, the fluctuations increase from
$\av{(\Delta\xbar)^2}=0$ at $t\to 0$ to their asymptotic value at $t\to \infty$, $\x2th(1-\al)$.
However, as can be seen also in Eq. \eqref{xbar_avg_longt}, for $\al>1/3$ the fluctuations display
a maximum and decay to their asymptotic limit from above. We found numerically that the value of
the maximal fluctuations scales roughly as $\al^{-1/2}$ (not shown). It can also be seen that for
almost all times and all values of $\alpha$, the fluctuations $\av{(\Delta\xbar)^2}$ decrease as
the diffusion becomes more ``normal'' (increasing $\alpha$), as expected. However, this pattern
surprisingly breaks down for $\alpha \lesssim 0.15$, for which there is a time window when the
fluctuations increase with $\alpha$.

It is straightforward to generalize our results to any initial condition with first moment
$\av{x_0}$ and second moment $\av{x_0^2}$. The first moment of the time-average is
$\av{\xbar}=\av{x_0}E_{\al,2}\left[-(t/\tau)^{\alpha}\right]$, which decays for long-times as
$\av{\xbar}\sim \frac{\av{x_0}}{\Gamma(2-\al)}\left(\frac{t}{\tau}\right)^{-\al}$.
The second moment is
\begin{align}
\label{timeavg_fluc_mittag_allx0} &\av{\xbar^2}=(1-\al)\x2th+2\alpha
\left[2\x2th-\av{x_0^2}\right]E_{\alpha,3}\left[-(t/\tau)^{\alpha}\right]\nonumber\\&+2(1+\al)\left[\av{x_0^2}-\x2th\right]E_{\alpha,3}\left[-2(t/\tau)^{\alpha}\right],
\end{align}
from which the fluctuations directly follow. For long times,
\begin{align*}
\label{xbar_avg_longt_allx0} \av{(\Delta\overline{x})^2}&\approx (1-\alpha)\x2th
\nonumber\\&+\frac{(3\alpha-1)\x2th+(1-\al)\av{x_0^2}}
{\Gamma(3-\alpha)}\left(\frac{t}{\tau}\right)^{-\alpha}.
\end{align*}
For short times,
\begin{equation*}
\label{xbar_avg_shortt_allx0} \av{(\Delta\overline{x})^2}\approx \av{(\Delta x_0)^2} -
2\left[\frac{\av{(\Delta
x_0)^2}}{\Gamma(2+\al)}-\frac{2\x2th}{\Gamma(3+\al)}\right]\left(\frac{t}{\tau}\right)^{\al},
\end{equation*}
where $\av{(\Delta x_0)^2}=\av{x_0^2}-\av{x_0}^2$. According to the last two equations, if the
system is already in equilibrium at $t=0$ such that $\av{x_0^2}=\x2th$, the fluctuations
monotonically decay, for all $\alpha$, from $\x2th$ at $t=0$ to $\x2th(1-\al)$ at $t\to \infty$.

For $\al=1$ (and $x_0=0$), we find the known result \cite{Risken}:
\begin{equation}
\label{xbar_fluc_alpha1}
\av{(\Delta\xbar)^2}_{\alpha=1}=\left(\frac{\tau}{t}\right)^{2}\left(4e^{-t/\tau}-e^{-2t/\tau}+\frac{2t}{\tau}-3\right).
\end{equation}
To derive the last equation, we used the relation $E_{1,3}(z)=[e^z-z-1]/z^2$. Since the ordinary
($\al=1$) Ornstein-Uhlenbeck process is a Gaussian process \cite{VanKampen}, the PDF of $\xbar$ is
a Gaussian too, with the variance indicated by Eq. \eqref{xbar_fluc_alpha1}.

\subsubsection{Fractional Kramers equation}

Finally, we remark on the connection between the fractional Feynman-Kac equation of this subsection
and an important class of processes in which the \emph{velocity} of the particle is the quantity
undergoing sub-diffusion. For such processes, Friedrich and coworkers have recently developed a
fractional Kramers equation for the joint position-velocity PDF
\cite{FriedrichPRL06,FriedrichPRE06}. For example, consider a Rayleigh-like model in which a free,
heavy test particle of mass $M$ collides with light bath particles at random times, but where the
times between collisions are distributed according to $\psi(\tau)\sim \tau^{-(1+\al)}$. The PDF of
the velocity of the test particle, $G(v,t)$, satisfies the fractional Fokker-Planck equation
\cite{BarkaiJPC00}:
\begin{equation*}
\frac{\di}{\di t}G(v,t)=\gamma_{\al} \left[\frac{k_{B}T}{M}\frac{\di^2}{\di v^2}+\frac{\di}{\di
v}v\right]{\cal D}_{\textrm{RL},t}^{1-\al}G(v,t),
\end{equation*}
where ${\cal D}_{\textrm{RL},t}^{1-\al}$ is the Riemann-Liouville fractional derivative operator
(see Section \ref{sect_specialcases}) and $\gamma_{\al}$ is the damping coefficient. Since in the
collisions model $x(t)=\int_0^t v(\tau)d\tau$, $x$ is a functional of the trajectory $v(\tau)$, and
therefore, the joint PDF of $x$ and $v$, $G(v,x,t)$, is described by our fractional Feynman-Kac
equation. Denoting the Fourier transform $x\to p$ of $G(v,x,t)$ as $G(v,p,t)$, we have (see Eq.
\eqref{forward_eq_fourier}),
\begin{align}
\label{fractional_kramers} \frac{\di}{\di t}G(v,p,t)&=ipvG(v,p,t)\\&+\gamma_{\al}
\left[\frac{k_{B}T}{M}\frac{\di^2}{\di v^2}+\frac{\di}{\di v}v\right]{\cal
D}_{t}^{1-\al}G(v,p,t)\nonumber,
\end{align}
where ${\cal D}_{t}^{1-\al}$ is the fractional substantial derivative, here equal in Laplace $s$
space to $(s-ipv)^{1-\al}$. Within this model, for $0<\al<1$ the motion is ballistic, $\av{x^2}\sim
t^2$, while for $\al=1$ it is diffusive, $\av{x^2}\sim t$ (see Eq. \eqref{xbar_avg_longt}). Eq.
\eqref{fractional_kramers} is exactly equal to the fractional Kramers equation derived by Friedrich
and coworkers \cite{FriedrichPRL06,FriedrichPRE06}, and in that sense, our results generalize their
pioneering work.


\section{Summary and discussion}

Time-averages of sub-diffusive continuous-time random walks (CTRW) in binding fields are known to
exhibit weak ergodicity breaking and were thus the subject of recent interest. In this paper, we
used the Feynman-Kac approach to develop a general equation for time-averages of CTRW (Eq.
\eqref{forward_eq_derive}), which can be seen as a fractional generalization of the Feynman-Kac
equation for Brownian motion. The equation we derived describes the distribution of time-averages
for all observables, potentials, and times. We also derived a backward equation (Eq.
\eqref{backward_eq}) which is useful in practical problems.

We investigated two applications of our equations: the occupation fraction in the positive half of
a box, and the time-averaged position in a harmonic potential. In both cases, we obtained
expressions for the PDF for long times and for short times and calculated the fluctuations. We
found that the fluctuations decay as $t^{-\al}$ to their asymptotic limit, which is non-zero for
anomalous diffusion, $\al<1$. Our fractional Feynman-Kac equation thus provides a general tool for
the treatment of time-averages and for the study of the kinetics of weak ergodicity breaking.


Recently, the occupation time functional has been studied in the context of dynamical systems with
an infinite (non-normalizable) invariant measure \cite{Zweimuller}. It remains to be seen whether a
framework similar to that of the fractional Feynman-Kac equation could be developed for general
functionals of these processes. We also note that while the (integer) Feynman-Kac equation can be
derived using path integrals \cite{MajumdarReview}, a path integral approach for functionals of
anomalous sub-diffusion is still awaiting (but see preliminary results in the upcoming book
\cite{FractionalDynamicsBook}).


\section*{Acknowledgements}

We thank David Kessler and Lior Turgeman from Bar-Ilan University for discussions and the Israel
Science Foundation for financial support. S.C. thanks Erez Levanon from Bar-Ilan University for his
hospitality during the course of this project.

\section*{Appendix A: Time-dependent forces}

In our model of CTRW with a time-dependent force, jump probabilities are determined according to
the force at the time of the jump. To derive an equation for $G(x,A,t)$ in that case, we rewrite
Eq. \eqref{Q_recursion} as follows:
\begin{align}
&\chi(x,A,t)=G_0(x)\delta(A)\delta(t)\\
&+\int_0^t\psi(\tau)L(x+a,t)\chi[x+a,A-\tau U(x+a),t-\tau]d\tau \nonumber
\\&+\int_0^t\psi(\tau)R(x-a,t)\chi[x-a,A-\tau U(x-a),t-\tau]d\tau \nonumber.
\end{align}
Note that the jump probabilities are time-dependent (but have no memory). Laplace transforming
$A\to p$ and $t\to s$, using the Laplace identity ${\cal L}\{tf(t)\}=\mds f(s)$,
\begin{align*}
&\chi(x,p,s)=G_0(x)\\
&+L\left(x+a,\mds\right)\hat\psi[s+pU(x+a)]\chi(x+a,p,s) \nonumber \\
&+R\left(x-a,\mds\right)\hat\psi[s+pU(x-a)]\chi(x-a,p,s) \nonumber.
\end{align*}
Fourier transforming $x\to k$,
\begin{align*}
\chi(k,p,s)=\hat{G}_0(k)+& \left[\cos(ka)+i\sin(ka)\frac{aF\left(-i\frac{\di}{\di
k},\mds\right)}{2k_{B}T}\right]\times\\ &\hat\psi\left[s+pU\left(-i\frac{\di}{\di
k}\right)\right]\chi(k,p,s)\nonumber.
\end{align*}
Continuing as in Section \ref{sect_forward_derivation}, we find the formal solutions for
$\chi(k,p,s)$ and $G(k,p,s)$ and then take the continuum limit. This gives:
\begin{align*}
&sG(k,p,s)-\hat{G}_0(k)=-pU\left(-i\frac{\di}{\di
k}\right)G(k,p,s)\\&-\Ka\left[k^2-ik\frac{F\left(-i\frac{\di}{\di
k},\mds\right)}{k_{B}T}\right]\times\nonumber \\&\left[s+pU\left(-i\frac{\di}{\di
k}\right)\right]^{1-\alpha}G(k,p,s)\nonumber.
\end{align*}
Inverting $k\to x, s\to t$, we obtain the fractional Feynman-Kac equation for a time-dependent
force:
\begin{equation}
\frac{\di}{\di t}G(x,p,t)=\Ka\Lfpt{\cal D}_t^{1-\alpha}G(x,p,t)-pU(x)G(x,p,t),
\end{equation}
where
\begin{equation*}
\Lfpt=\frac{\di^2}{\di x^2}-\frac{\di}{\di x}\frac{F(x,t)}{k_{B}T}
\end{equation*}
is the time-dependent Fokker-Planck operator.

\section*{Appendix B: The simulation method}

The fractional Feynman-Kac equation describes the joint PDF of $x$ and $A$ in the continuum limit
of CTRW. In this limit, $a\to 0$ and $B_{\alpha}\to 0$ but the generalized diffusion coefficient
$K_{\alpha}=a^2/(2B_{\alpha})$ (Eq. \eqref{gen_diff_coeff}) is kept finite \cite{BarkaiPRE00}. We
simulate trajectories of this process as follows \cite{HanggiSimulations}. We place a particle on a
one-dimensional lattice in initial position $x_0$, where usually $x_0=0$. We set the lattice
spacing $a$ and the generalized diffusion coefficient $\Ka$ and determine $B_{\alpha}=a^2/(2\Ka)$.
Waiting times are then drawn for $\alpha=1$ from an exponential distribution
$\psi(\tau)=e^{-\tau/\tau_0}/\tau_0$ with mean $\tau_0=B_1$. This is implemented by setting
$\tau=-\tau_0\ln(u)$, where $u$ is a number uniformly distributed in $[0,1]$. For $\alpha<1$, we
set $\tau_0=[B_{\alpha}/\Gamma(1-\alpha)]^{1/\alpha}$ and $\tau=\tau_0u^{-1/\alpha}$, which
corresponds to $\psi(\tau)=\frac{B_{\alpha}}{|\Gamma(-\alpha)|}\tau^{-(1+\alpha)}$ ($\tau \geq
\tau_0$; see Eq. \eqref{eq_pdf_psi}). After waiting time $\tau$, we move the particle right or left
with probabilities $R(x)$ or $L(x)$, respectively, as given by Eq. \eqref{left_right_force}. For
the harmonic potential, Eq. \eqref{left_right_force} gives
$R(x)=\frac{1}{2}\left(1-\frac{ax}{2\x2th}\right)$ and
$L(x)=\frac{1}{2}\left(1+\frac{ax}{2\x2th}\right)$. Since the typical $x$ is of the order of
$\sqrt{\x2th}$, it is sufficient to choose $a\ll \sqrt{\x2th}$ to guarantee that $0<R(x),L(x)<1$
(see discussion in \cite{BarkaiPRE06}). For the box, $R(x)=L(x)=1/2$ and we make the boundaries at
$x=\pm\frac{L}{2}$ reflecting.

The parameters we used in the simulations were as follows. In all simulations, we used $a=0.1$ or
smaller, and each curve represents at least $10^4$ trajectories. For the occupation time in a box,
we set $L=2$ and $\Ka=1$, and the final simulation time in Figure \ref{Fig_occ_PDF} was $t=10^3$.
For the time-averaged position in the harmonic potential, we set $\Ka=1/2$ and $\x2th=1/2$ (or
$\tau^{\al}=1$). In Figure \ref{Fig_xbar_PDF}, the final simulation times were as follows. For the
long-times limit (top panel) we used $t=10^7, 10^4, 10^3, 10^3$ for $\al=0.25,0.5,0.75,1$,
respectively. For the short times (bottom panel), we used $t=10^{-3},10^{-2},10^{-1}$ for $\al=1$,
$t=10^{-5},10^{-4},10^{-3}$ for $\al=0.5$, and $t=10^{-6},10^{-5},10^{-4}$ for $\al=0.25$.

\section*{Appendix C: The $t\to \infty$ distribution of the time-averaged position in a harmonic potential.}

Consider the time-averaged position, $\xbar=\frac{1}{t}\int_0^t x(\tau)d\tau$, for a sub-diffusing
particle in a harmonic potential, $V(x)=m\omega^2x^2/2$. Using the thermal second moment,
$\x2th=k_BT/(m\omega^2)$, and for $t\to \infty$, we have
\begin{equation*}
G(\xbar)=\frac{1}{\sqrt{\x2th}}g\left(\frac{\xbar}{\sqrt{\x2th}}\right),
\end{equation*}
where
\begin{widetext}
\begin{align}
\label{G_stationary_xbar_full} &g(y)=\frac{\sin(\pi\al)}{\pi}\times\br \left\{e^{y^2/2} y
\Gamma\left(\frac{\al}{2}\right)
\Gamma(1+\al)\left[\M\left(\frac{1-\al}{2},\frac{1}{2},-\frac{y^2}{2}\right)
\U\left(1+\frac{\al}{2},\frac{3}{2},\frac{y^2}{2}\right)+2\M\left(\frac{1-\al}{2},\frac{3}{2},-\frac{y^2}{2}\right)
\U\left(\frac{\al}{2},\frac{1}{2},\frac{y^2}{2}\right)\right]\right.\br\left.+\sqrt{2}\Gamma(\al)
\Gamma\left(\frac{1+\al}{2}\right)\left[y^2 \al
\M\left(\frac{1+\al}{2},\frac{3}{2},\frac{y^2}{2}\right)
\U\left(1+\frac{\al}{2},\frac{3}{2},\frac{y^2}{2}\right)+2\M\left(\frac{1+\al}{2},\frac{1}{2},\frac{y^2}{2}\right)
\U\left(\frac{\al}{2},\frac{1}{2},\frac{y^2}{2}\right)\right] \right\}\times \br \left\{2^{2+\al}
y^2 \Gamma^2\left(1+\frac{\al}{2}\right) \left[e^{y^2}
\M^2\left(\frac{1-\al}{2},\frac{3}{2},-\frac{y^2}{2}\right)-2 \cos(\pi\al)
\M^2\left(1+\frac{\al}{2},\frac{3}{2},\frac{y^2}{2}\right)\right]\right.\br+4\sqrt{2}e^{y^2}\sqrt{\pi}y
\Gamma(1+\al)\M\left(\frac{1-\al}{2},\frac{3}{2},-\frac{y^2}{2}\right)
\M\left(-\frac{\al}{2},\frac{1}{2},-\frac{y^2}{2}\right)\br+2^{1+\al}
\Gamma^2\left(\frac{1+\al}{2}\right) \left[e^{y^2}
\M^2\left(-\frac{\al}{2},\frac{1}{2},-\frac{y^2}{2}\right)+2\cos(\pi\al)
\M^2\left(\frac{1+\al}{2},\frac{1}{2},\frac{y^2}{2}\right)\right]\br\left.+2^{-\al} y^2
\Gamma^2(1+\al) \U^2\left(1+\frac{\al}{2},\frac{3}{2},\frac{y^2}{2}\right)\right\}^{-1}.
\end{align}
\end{widetext}
In the last equation, $\M(a,b,z)$ is the confluent hypergeometric (or Kummer's) function of the
first kind and $\U(a,b,z)$ is the confluent hypergeometric (or Kummer's) function of the second
kind \cite{Abramowitz}. Eq. \eqref{G_stationary_xbar_full} is valid for $y>0$. Due to the symmetry
of the potential, $g(-y)=g(y)$.

\bibliographystyle{unsrt}
\bibliography{../functionals}

\begin{thebibliography}{10}

\bibitem{Havlin}
S.~Havlin and D.~{ben-Avraham}.
\newblock Diffusion in disordered media.
\newblock {\em Adv. Phys.}, 36:695, 1987.

\bibitem{Bouchaud}
J.~P. Bouchaud and A.~Georges.
\newblock Anomalous diffusion in disordered media: Statistical mechanisms,
  models and physical applications.
\newblock {\em Phys. Rep.}, 195:127, 1990.

\bibitem{KlafterReview2000}
R.~Metzler and J.~Klafter.
\newblock The random walk's guide to anomalous diffusion: A fractional dynamics
  approach.
\newblock {\em Phys. Rep.}, 339:1, 2000.

\bibitem{AnomalousTransportBook}
R.~Klages, G.~Radons, and I.~M. Sokolov, editors.
\newblock {\em Anomalous Transport: Foundations and Applications}.
\newblock Wiley-VCH, Weinheim, 2008.

\bibitem{MontrollWeiss}
E.~W. Montroll and G.~H. Weiss.
\newblock Random walks on lattices{. II}.
\newblock {\em J. Math. Phys.}, 6:167, 1965.

\bibitem{ScherMontroll}
H.~Scher and E.~Montroll.
\newblock Anomalous transit-time dispersion in amorphous solids.
\newblock {\em Phys. Rev. B}, 12:2455, 1975.

\bibitem{BarkaiPRL05}
G.~Bel and E.~Barkai.
\newblock Weak ergodicity breaking in the continuous-time random walk.
\newblock {\em Phys. Rev. Lett.}, 94:240602, 2005.

\bibitem{BarkaiJSP06}
E.~Barkai.
\newblock Residence time statistics for normal and fractional diffusion in a
  force field.
\newblock {\em J. Stat. Phys.}, 123:883, 2006.

\bibitem{BarkaiPRL07}
A.~Rebenshtok and E.~Barkai.
\newblock Distribution of time-averaged observables for weak ergodicity
  breaking.
\newblock {\em Phys. Rev. Lett.}, 99:210601, 2007.

\bibitem{BarkaiJSP08}
A.~Rebenshtok and E.~Barkai.
\newblock Weakly non-ergodic statistical physics.
\newblock {\em J. Stat. Phys.}, 133:565, 2008.

\bibitem{BouchaudWEB}
J.~P. Bouchaud.
\newblock Weak ergodicity breaking and aging in disordered systems.
\newblock {\em J. de Physique I}, 2:1705, 1992.

\bibitem{GrebenkovPRE}
D.~S. Grebenkov.
\newblock Residence times and other functionals of reflected {Brownian} motion.
\newblock {\em Phys. Rev. E}, 76:041139, 2007.

\bibitem{OccTimeMajumdarPRL}
S.~N. Majumdar and A.~Comtet.
\newblock Local and occupation time of a particle diffusing in a random medium.
\newblock {\em Phys. Rev. Lett.}, 89:060601, 2002.

\bibitem{OccupationExeperimental}
G.~Zumofen, J.~Hohlbein, and C.~G. H\"{u}bner.
\newblock Recurrence and photon statistics in fluorescence fluctuation
  spectroscopy.
\newblock {\em Phys. Rev. Lett.}, 93:260601, 2004.

\bibitem{AgmonResidenceMoments}
N.~Agmon.
\newblock The residence time equation.
\newblock {\em Chem. Phys. Lett.}, 497:184, 2010.

\bibitem{AnomalousBeads1}
I.~Y. Wong, M.~L. Gardel, D.~R. Reichman, E.~R. Weeks, M.~T. Valentine, A.~R.
  Bausch, and D.~A. Weitz.
\newblock Anomalous diffusion probes microstructure dynamics of entangled
  {F}-actin networks.
\newblock {\em Phys. Rev. Lett.}, 92:178101, 2004.

\bibitem{AnomalousBeads2}
G.~Pesce, L.~Selvaggi, A.~Caporali, A.~C.~De Luca, A.~Puppo, G.~Rusciano, and
  A~Sasso.
\newblock Mechanical changes of living oocytes at maturation investigated by
  multiple particle tracking.
\newblock {\em Appl. Phys. Lett.}, 95:093702, 2009.

\bibitem{AnomalousDNA}
G.~G. Cabal, A.~Genovesio, S.~Rodriguez-Navarro, C.~Zimmer, O.~Gadal, A.~Lesne,
  H.~Buc, F.~Feuerbach-Fournier, J.-C. Olivo-Marin, E.~C. Hurt, and
  U.~Nehrbass.
\newblock {SAGA} interacting factors confine sub-diffusion of transcribed genes
  to the nuclear envelope.
\newblock {\em Nature}, 441:770, 2006.

\bibitem{AnomalousTelomere}
I.~Bronstein, Y.~Israel, E.~Kepten, S.~Mai, Y.~Shav-Tal, E.~Barkai, and
  Y.~Garini.
\newblock Transient anomalous diffusion of telomeres in the nucleus of
  mammalian cells.
\newblock {\em Phys. Rev. Lett.}, 103:018102, 2009.

\bibitem{AnomalousmRNA}
I.~Golding and E.~C. Cox.
\newblock Physical nature of bacterial cytoplasm.
\newblock {\em Phys. Rev. Lett.}, 96:098102, 2006.

\bibitem{AnomalousLipid1}
I.~M. Toli\'{c}-N{\o}rrelykke, E.-L. Munteanu, G.~Thon, L.~Oddershede, and
  K.~Berg-S{\o}rensen.
\newblock Anomalous diffusion in living yeast cells.
\newblock {\em Phys. Rev. Lett.}, 93:078102, 2004.

\bibitem{MajumdarReview}
S.~N. Majumdar.
\newblock {Brownian} functionals in physics and computer science.
\newblock {\em Curr. Sci.}, 89:2076, 2005.

\bibitem{Kac1949}
M.~Kac.
\newblock On distributions of certain {Wiener} functionals.
\newblock {\em Trans. Am. Math. Soc.}, 65:1, 1949.

\bibitem{BarkaiPRL09}
L.~Turgeman, S.~Carmi, and E.~Barkai.
\newblock Fractional {Feynman-Kac} equation for non-{Brownian} functionals.
\newblock {\em Phys. Rev. Lett.}, 103:190201, 2009.

\bibitem{BarkaiJSP10}
S.~Carmi, L.~Turgeman, and E.~Barkai.
\newblock On distributions of functionals of anomalous diffusion paths.
\newblock {\em J. Stat. Phys.}, 141:1071, 2010.

\bibitem{FriedrichPRL06}
R.~Friedrich, F.~Jenko, A.~Baule, and S.~Eule.
\newblock Anomalous diffusion of inertial, weakly damped particles.
\newblock {\em Phys. Rev. Lett.}, 96:230601, 2006.

\bibitem{FriedrichPRE06}
R.~Friedrich, F.~Jenko, A.~Baule, and S.~Eule.
\newblock Exact solution of a generalized {Kramers-Fokker-Planck} equation
  retaining retardation effects.
\newblock {\em Phys. Rev. E}, 74:041103, 2006.

\bibitem{BarkaiPRE00}
E.~Barkai, R.~Metzler, and J.~Klafter.
\newblock From continuous time random walks to the fractional {Fokker-Planck}
  equation.
\newblock {\em Phys. Rev. E}, 61:132, 2000.

\bibitem{WaitingTimeCutoff}
T.~Miyaguchi and T.~Akimoto.
\newblock Ultraslow convergence to ergodicity in transient subdiffusion.
\newblock {\em Phys. Rev. E}, 83:062101, 2011.

\bibitem{MetzlerLevyWalk}
I.~M. Sokolov and R.~Metzler.
\newblock Towards deterministic equations for {L\'{e}vy} walks: {The}
  fractional material derivative.
\newblock {\em Phys. Rev. E}, 67:010101(R), 2003.

\bibitem{BarkaiPRL99}
R.~Metzler, E.~Barkai, and J.~Klafter.
\newblock Anomalous diffusion and relaxation close to thermal equilibrium: A
  fractional {Fokker-Planck} equation approach.
\newblock {\em Phys. Rev. Lett.}, 82:3563, 1999.

\bibitem{BarkaiPRE01}
E.~Barkai.
\newblock Fractional {Fokker-Planck} equation, solution, and application.
\newblock {\em Phys. Rev. E}, 63:046118, 2001.

\bibitem{TimeDependentSokolov}
I.~M. Sokolov and J.~Klafter.
\newblock Field-induced dispersion in subdiffusion.
\newblock {\em Phys. Rev. Lett.}, 97:140602, 2006.

\bibitem{TimeDependentHanggi}
E.~Heinsalu, M.~Patriarca, I.~Goychuk, and P.~H\"{a}nggi.
\newblock Use and abuse of a fractional {F}okker-{P}lanck dynamics for
  time-dependent driving.
\newblock {\em Phys. Rev. Lett.}, 99:120602, 2007.

\bibitem{TimeDependentWeron}
M.~Magdziarz and A.~Weron.
\newblock Equivalence of the fractional {F}okker-{P}lanck and subordinated
  {L}angevin equations: The case of a time-dependent force.
\newblock {\em Phys. Rev. Lett.}, 101:210601, 2008.

\bibitem{TimeDependentHenry}
B.~I. Henry, T.~A.~M. Langlands, and P.~Straka.
\newblock Fractional {F}okker-{P}lanck equations for subdiffusion with space-
  and time-dependent forces.
\newblock {\em Phys. Rev. Lett.}, 105:170602, 2010.

\bibitem{FriedrichEPL09}
S.~Eule and R.~Friedrich.
\newblock Subordinated {Langevin} equations for anomalous diffusion in external
  potentials--- {Biasing} and decoupled external forces.
\newblock {\em EPL}, 86:30008, 2009.

\bibitem{GodrecheLuck}
C.~Godr\`{e}che and J.~M. Luck.
\newblock Statistics of the occupation time of renewal processes.
\newblock {\em J. Stat. Phys.}, 104:489, 2001.

\bibitem{Lamperti}
J.~Lamperti.
\newblock An occupation time theorem for a class of stochastic processes.
\newblock {\em Trans. Am. Math. Soc.}, 88:380, 1958.

\bibitem{MargolinJSP}
G.~Margolin and E.~Barkai.
\newblock Non-ergodicity of a time series obeying {L}\'{e}vy statistics.
\newblock {\em J. Stat. Phys.}, 122:137, 2006.

\bibitem{Redner_book}
S.~Redner.
\newblock {\em A Guide to First-Passage Processes}.
\newblock Cambridge University Press, 2001.

\bibitem{Kac1951}
M.~Kac.
\newblock On some connections between probability theory and differential and
  integral equations.
\newblock In {\em Second Berkeley Symposium on Mathematical Statistics and
  Probability}, page 189, Berkeley, CA, USA, 1951. University of California
  Press.

\bibitem{BarkaiPRE06}
G.~Bel and E.~Barkai.
\newblock Random walk to a nonergodic equilibrium concept.
\newblock {\em Phys. Rev. E}, 73:016125, 2006.

\bibitem{BarkaiJPC00}
E.~Barkai and R.~J. Silbey.
\newblock Fractional {K}ramers equation.
\newblock {\em J. Phys. Chem. B}, 104:3866, 2000.

\bibitem{Podlubny}
I.~Podlubny.
\newblock {\em Fractional Differential Equations}.
\newblock Academic Press, New York, 1999.

\bibitem{PodlubnyMLF}
I.~Podlubny.
\newblock http://www.mathworks.com/matlabcentral, 2009.

\bibitem{Risken}
H.~Risken.
\newblock {\em The {Fokker}–{Planck} Equation: Methods of Solution and
  Applications}.
\newblock Springer, 2nd edition, 1989.

\bibitem{VanKampen}
N.~G.~Van Kampen.
\newblock {\em Stochastic Processes in Physics and Chemistry}.
\newblock North Holland, 3rd edition, 2007.

\bibitem{Zweimuller}
M.~Thaler and R.~Zweim\"{u}ller.
\newblock Distributional limit theorems in infinite ergodic theory.
\newblock {\em Probab. Theory Rel.}, 135:15, 2006.

\bibitem{FractionalDynamicsBook}
Fractional dynamics: Recent advances.
\newblock 2011.

\bibitem{HanggiSimulations}
E.~Heinsalu, M.~Patriarca, I.~Goychuk, G.~Schmid, and P.~H\"{a}nggi.
\newblock Fractional {Fokker-Planck} dynamics: {Numerical} algorithm and
  simulations.
\newblock {\em Phys. Rev. E}, 73:046133, 2006.

\bibitem{Abramowitz}
M.~Abramowitz and I.~A. Stegun, editors.
\newblock {\em Handbook of Mathematical Functions with Formulas, Graphs, and
  Mathematical Tables}.
\newblock Dover Publications, New York, 1972.

\end{thebibliography}

\end{document}